\documentclass[a4paper,12pt]{article} 
\usepackage[utf8]{inputenc} 
\usepackage[english]{babel}
\usepackage [autostyle, english = american]{csquotes}
\usepackage{graphicx}
\MakeOuterQuote{"}
\usepackage{graphicx} 
\usepackage{caption}
\usepackage{subcaption}
\usepackage{float} 
\usepackage{tabularx}
\usepackage[round,authoryear]{natbib}
\usepackage{ragged2e}
\usepackage{booktabs}   
\usepackage{caption}    
\usepackage{graphicx}   
\usepackage{array}      
\usepackage{booktabs} 
\usepackage{threeparttable} 
\usepackage{siunitx}
\usepackage{graphicx}  
\usepackage{longtable}  
\usepackage{amsmath}    
\usepackage{booktabs}
\usepackage{geometry}

\usepackage{longtable}
\usepackage{array}

\date{\today}

\begin{document}
\begin{titlepage}
\title{Long-term health and human capital effects of massive investments in public health and education: Evidence from Cuba\thanks{The authors gratefully acknowledge helpful comments from Thomas Barnebeck Andersen, Gregory Clark, Matthew Curtis, Seetha Menon, Paul Sharp, and Anthony Wray, as well as from participants at IEEE 2025, and EALE 2025.}}

\author{Giovanni Mellace\footnote{Department of Economics, Campusvej 55, 5230 Odense M, Denmark, giome@sam.sdu.dk} \\
{\small University of Southern Denmark}
\\
\and
Rok Spruk\footnote{School of Economics and Business, Kardeljeva ploščad 17,
1000 Ljubljana, Slovenia, rok.spruk@ef.uni-lj.si} \\
{\small  University of Ljubljana}
  \\}
  
\maketitle
\thispagestyle{empty}
\begin{abstract}
We estimate long-run effects of Cuba's 1961 National Health Service and contemporaneous National Literacy Campaign using synthetic-control methods on newly assembled series for 21 former European colonies in the Americas, 1900-2022. Relative to synthetic Cuba, infant mortality falls 15--29 percent and average years of schooling rise 1.5-2 years; both effects are large, persistent, and robust to augmented SCM, synthetic difference-in-differences, interactive fixed effects, and matrix completion. Life-expectancy gains attenuate after 1990, consistent with the post-Soviet Special Period, suggesting that bundled health and literacy reforms permanently raise early-life survival and human capital, with smaller and less robust effects on adult longevity.

\end{abstract}

\end{titlepage}
\newpage 
\section{Introduction}
Investments in public health services are fundamental to enhancing population health outcomes and fostering human capital development. The implementation of universal health care systems has been associated with significant improvements in infant mortality rates, life expectancy, and educational attainment \citep{grossman1972concept}. Understanding the long-term effects of such systems, especially in varying socio-economic contexts, is crucial for informing policy decisions.

In 1960, Cuba established a comprehensive National Health Service aimed at providing equitable and accessible health care to all citizens \citep{keck2012curious}. This significant policy shift transformed Cuba's health care landscape and presents a unique opportunity to examine the causal impact of large-scale public health investments on key health and human capital indicators. While previous research has highlighted Cuba's notable health achievements despite economic challenges \citep{nikelly1988,feinsilver2023healing}, there is a lack of empirical studies quantifying the long-term effects of its National Health Service on public health outcomes and educational attainment.

This study examines the contribution of Cuba's National Health Service to public health and human capital formation from 1870 to 2022. By exploiting the establishment of the National Health Service as a source of variation in long-term public health and human capital formation, we aim to estimate the causal effects of universal health care provision. 

Our identification strategy utilizes the synthetic control method developed by \citet{abadie2003economic} and \citet{abadie2010synthetic} to build a synthetic version of Cuba that closely mirrors its pre-intervention trends as a weighted combination of former European colonies in the Western Hemisphere that did not implement comprehensive universal health care during the same period. Comparing Cuba's actual outcomes with this synthetic control allows us to estimate the causal effects of the National Health Service on infant mortality, life expectancy, and human capital formation.

Our findings indicate that the bundle of universal health-care and mass-literacy reforms is associated with a large and persistent reduction in infant mortality of 15--29 percent across estimators, a 1.5--2 year rise in average years of schooling, and a positive but more modest gain in life expectancy at birth that attenuates over time and becomes statistically imprecise under more flexible specifications. These effects survive an extensive battery of placebo, leave-one-out, in-time, and alternative-donor exercises, and remain when the donor pool is expanded to include 44 U.S.\ states with comparable pre-reform trajectories.

This study makes three contributions. \textit{First}, it provides one of the longest-horizon quasi-experimental estimates of the joint effect of universal health-care and mass-literacy reform in a developing-country setting, spanning more than six decades of post-treatment data. \textit{Second}, it documents a sharp heterogeneity in the persistence of treatment effects across outcomes: while infant mortality and schooling display durable, structurally meaningful divergences from synthetic controls, life expectancy is more shock-sensitive and weakens after the 1990 collapse of Soviet support. This pattern speaks to the longstanding debate about which dimensions of population health are most responsive to health-system inputs versus broader macroeconomic conditions. \textit{Third}, the paper assembles newly harmonized historical series on infant mortality, life expectancy, and schooling for 21 former European colonies in the Americas back to 1900 and links them to standardized post-1950 UN benchmarks, providing a reusable panel for comparable studies of long-run health reforms. As a complement to the baseline synthetic-control analysis, we report a battery of robustness checks based on augmented SCM with conformal inference \citep{chernozhukov2021exact}, synthetic difference-in-differences \citep{arkhangelsky2021synthetic}, interactive fixed effects \citep{xu2017generalized}, and matrix completion \citep{athey2021matrix}, all of which yield mutually consistent estimates for infant mortality and schooling and the same pattern of attenuation for life expectancy.

\paragraph{Related literature.} The paper connects to four literatures. Most directly, it builds on the empirical study of how early-life health interventions shape long-run human capital and economic outcomes. Historical evidence on hookworm eradication \citep{bleakley2010health}, the U.S.\ public-health transition \citep{cutler2005role}, fetal-origin shocks \citep{AlmondCurrie2011HumanCapital}, mass deworming \citep{miguel2004worms}, and the rollout of the Swedish welfare state \citep{bhalotra2022health} consistently finds that improvements in early-life conditions yield large, persistent gains in schooling, earnings, and adult health. Our setting extends this evidence to a developing-country context with universal coverage and a long post-reform horizon. A second strand concerns the joint determination of health and human capital in models of intergenerational investment \citep{grossman1972concept,BeckerTomes1979,BarroBecker1989,Heckman2007HumanCapital}. The Cuban case, in which a large health reform and a mass literacy campaign are implemented essentially simultaneously, provides an unusually clean window into the complementarity emphasized by these models.

A third strand is the political economy of state capacity. Recent work argues that the ability of states to deliver core public goods reflects, and reinforces, long-run institutional capacity \citep{BesleyPersson2009StateCapacity,DinceccoPrado2012FiscalCapacity,DinceccoKatz2016Capacity}, and that the consolidation of internal security is an important complement to service delivery \citep{AcemogluRobinson2006Institutions,AcemogluRobinsonSantos2013Violence}. Cuba's post-1959 reforms are a textbook case of rapid state-capacity expansion: hospitals, polyclinics, schools, and literacy brigades were embedded into previously peripheral rural areas. We interpret the durability of the infant-mortality and schooling gaps as evidence of this state-building project, while remaining cautious about attributing the entire effect to the health reform alone given the bundled nature of the post-revolutionary transformation. Our empirical strategy applies the synthetic control method of \citet{abadie2003economic,abadie2010synthetic,abadie2015comparative} and complements it with augmented synthetic control \citep{benmichael2021augmented,chernozhukov2021exact}, synthetic difference-in-differences \citep{arkhangelsky2021synthetic}, interactive fixed effects \citep{xu2017generalized}, and matrix completion \citep{athey2021matrix} as robustness checks.

The remainder of the paper is organized as follows. Section \ref{history} introduces the historical and institutional background. Section \ref{sec:theory} develops the theoretical mechanisms. Section \ref{data} discusses data and samples. Section \ref{identification} presents the identification strategy. Section \ref{sec:results} reports the results and robustness checks. Section \ref{sec:conclusion} concludes.

\section{Historical and Policy Background}\label{history}

This section provides a compact overview of Cuba's pre- and post-1959 health and education environment and motivates our identification strategy. The Cuban experience has long attracted attention as a case in which a low- to middle-income country attained public-health outcomes comparable to those of advanced economies despite limited income and an external embargo---the so-called Cuban health paradox \citep{SpiegelYassi2004Cuba,Evans2008Cuba}. A more detailed account, including the development of biotechnology and medical diplomacy and the legal and demographic changes of the post-1990 period, is provided in Appendix~\ref{app:history}.

\subsection{Pre-revolutionary Cuba}
On the eve of the 1959 revolution, Cuba's public-health landscape was characterized by deep urban--rural inequality. Despite scientific contributions in tropical medicine \citep{barnet1915finlay,chavescarballo2005finlay}, access to modern health care was concentrated in Havana and the principal cities and provided almost exclusively through private clinics \citep{danielson1979cuban,keck2012curious}. Rural areas had a single hospital and bore a heavy burden of tuberculosis, malaria, parasitic infections, and child malnutrition. Contemporary surveys document that only eleven percent of farm workers consumed milk and that rural infant mortality stood at roughly 100 per 1{,}000 live births \citep{truslow1951report}. Roughly 80 percent of children in the countryside had intestinal parasites and 60 percent of the population was chronically malnourished \citep{mcguire2005cuba,gorry2012cuba,delgadolegon2018cuba}. Pre-revolutionary life expectancy at birth has been estimated at 60--62 years in major cities and 45--50 years in rural areas, an urban--rural gap of 10--15 years. These inequalities were further reinforced by the widespread corruption and authoritarian repression of the Batista regime (1952--1959).

\subsection{The 1961 reforms: National Health Service and Literacy Campaign}
The post-1959 transformation centered on two contemporaneous, nation-wide reforms. First, the establishment of the \textit{Sistema Nacional de Salud} (National Health Service, NHS) in 1961 codified universal access to health care as a constitutional right, nationalized the hospital system, and rapidly extended primary care into rural areas through a polyclinic network \citep{conover1980cuba,boffey1978cuba,hirschfeld2017health}. Medical education was made free, mass-immunization programs were launched, and a nation-wide maternal and child program provided free prenatal care, childbirth, and infant vaccination \citep{bodenstein2010cuba}. \citet{DeVos2005CubaHealth} characterizes the system's subsequent development in four stages: foundation (1959--70), consolidation (1970--79), expansion of family medicine (1980--90), and reform (1990 onward).

Second, the National Literacy Campaign (\textit{Campa\~{n}a Nacional de Alfabetizaci\'{o}n}), also launched in 1961, mobilized more than 250{,}000 volunteer \textit{brigadistas} to teach reading and writing in rural and peri-urban communities \citep{leiner1987literacy,rey2021literacy}. Within eight months, the national illiteracy rate fell from roughly 20--40 percent to below 4 percent. Higher-education access was simultaneously expanded through gender quotas and other affirmative-action policies \citep{anderson1997che}.

\subsection{Bundled-reform interpretation}
The NHS and the Literacy Campaign were launched in the same year and operated, in many regions, through overlapping personnel and infrastructure. They were further embedded in a broader bundle of revolutionary changes, including land reform, nationalization of industries, large-scale elite emigration, the U.S.\ embargo from the early 1960s, and---through the 1980s---substantial Soviet subsidies that ended abruptly with the 1990--91 ``Special Period'' \citep{garfield1997impact,barry2000effect}. Our empirical strategy therefore identifies the joint effect of universal health-care provision and the simultaneous mass-literacy reform, rather than that of any single component. We return to this bundled-reform interpretation throughout the empirical analysis, in particular when interpreting the divergent post-1990 dynamics of life expectancy and the persistence of the schooling and infant-mortality gaps.

\section{Theoretical Mechanisms}\label{sec:theory}

The empirical analysis is guided by a simple conceptual framework that combines three building blocks: (i) the health-capital model in which health is a durable stock that responds to medical and nutritional investments \citep{grossman1972concept}; (ii) models of human-capital accumulation in which health, parental education, and schooling are jointly determined and mutually reinforcing \citep{BeckerTomes1979,BarroBecker1989,Heckman2007HumanCapital}; and (iii) the state-capacity literature linking the provision of public goods to administrative reach and the consolidation of internal security \citep{BesleyPersson2009StateCapacity,DinceccoPrado2012FiscalCapacity,DinceccoKatz2016Capacity,AcemogluRobinson2006Institutions,AcemogluRobinsonSantos2013Violence}.

These ingredients yield four testable implications that shape our empirical analysis. \textit{First}, the rollout of universal preventive and primary care into a high-mortality environment should generate a rapid and persistent decline in infant mortality and a slower-moving improvement in life expectancy at birth, with the strongest and most durable effects for outcomes tightly tied to early-life conditions. Existing evidence on parasitic-disease eradication, the U.S.\ public-health transition, fetal-origin shocks, and welfare-state expansion supports this prediction \citep{bleakley2010health,cutler2005role,AlmondCurrie2011HumanCapital,bhalotra2022health}. \textit{Second}, improvements in early-life survival and child health raise the expected returns to schooling, while the simultaneous removal of basic literacy constraints amplifies the complementarity between health and education \citep{miguel2004worms,bleakley2010health}. We therefore expect a gradual but enduring divergence in average years of schooling, larger and more persistent than the level effect of the literacy campaign on its own. \textit{Third}, the centrally directed expansion of clinics, schools, and literacy brigades constitutes a rapid increase in state capacity, lowering the marginal cost of subsequent public-goods provision and supporting long-run persistence of the gains. \textit{Fourth}, because life expectancy at birth is shaped by adult mortality, behavioral factors, and macroeconomic conditions in addition to health-system inputs, its long-run trajectory should be more exposed to subsequent shocks---such as the 1990 collapse of Soviet support and the U.S.\ embargo \citep{garfield1997impact,barry2000effect}---than infant mortality or average years of schooling.

In the empirical analysis that follows, we examine whether Cuba's experience after 1961 matches this set of predictions when compared to a carefully constructed synthetic control of similar countries (and, in robustness analyses, U.S.\ states) that did not implement comparable reforms. A more detailed development of the four mechanisms, including extensions to cohort persistence and intergenerational transmission, is provided in Appendix~\ref{app:theory}.

\section{Data}\label{data}

\subsection{Outcomes and samples}

Our sample comprises 21 former European colonies  in North and South America for the period 1870-2022. Using the average years of education as a rough proxy for human capital formation, we use \citet{BarroLee2010Education} estimates of the mean years of schooling across the full spectrum of age cohorts which enables a straightforward between-country and within-country comparison. The trajectories of infant mortality and life expectancy for the period 1900-2022 are constructed in two-fold step. In first step, we use the outcome trajectories for the period 1950-2022 from UN World Population Prospects as the underlying benchmark. In the second step, we link the benchmark series to the pre-1950 period by making use of \citet{Mitchell2010} estimates of infant mortality for the former European colonies in the Americas, extended back to 1900. In addition, the trajectories of life expectancy at birth from UN World Population Prospects for the period 1950-2022 are linked to the historical estimates of life expectancy for the same underlying sample for the period 1902-1950 based on the ground-breaking historical reconstruction of life expectancy estimates from \citet{Riley2005}, \citet{barbieri2015data}, \citet{Zijdeman2015}, and \citet{aburto2020dynamics} among several others.

\bigskip

We consider three specific outcomes to evaluate the long-term effectiveness of the establishment of National Health Service on the trajectories of public health, namely, infant mortality, life expectancy at birth and average years of education. First, infant mortality rate is defined as the number of deaths of infants under one year per 1,000 live births in a given year and is widely considered as perhaps one of the most comprehensive indicators of the overall health system performance that is also reflective of the quality of healthcare, maternal health, quality of nutrition and level of socioeconomic conditions. One of the key limitations behind the reconstruction of the infant mortality trajectory arises from the incompleteness of death registries in Cuba in early 1900s, as shown by \citet{DiazBriquets1983Mortality}. Therefore, the only plausible alternative is to rely on census data to reconstruct a plausible historical trajectory of infant mortality. For the period 1900-1904, we combine \citet{gonzalezquinones1996census} census-based estimates and \citet{collver1965birth} historical estimates for Latin American countries which provide a comprehensive and complete coverage of infant deaths from the birth registers. Since Cuba adopted the World Health Organization-based definitions of live birth, infant death and fetal deaths in 1965 \citep{catasus1977mortality, riosmassabot1983infant}, the infant mortality, linking pre-1950 series with the post-1950 benchmark provides a plausible long-term trajectory of infant morality rate that can be compared both across space and time.

\bigskip

Second, the reconstruction of the historical life expectancy trajectory consists of several steps. In the first step, we use \citet{AstorgaFitzgerald1998} estimates of life expectancy for the year 1900 which is around 32 years, as also reported by \citet{mcguire2005cuba}, and is consistent with the level calculated by Cuban government in 1975 which estimated life expectancy at birth in 1900 at 33.2 years \citep{centroestudios1976, farnos1977life}. We construct the life expectancy series for the period 1900-1950 by combining year-specific estimates from the original tables of \citet{alvarez1975life}, \citet{centroestudios1976}, \citet{mezquita1970life}, and \citet{debasa1971population} as well as \citet{AstorgaFitzgerald1998} and \citet{perezbrignoli2010demografia}. Our approach is simple and parsimonious as we compute the median estimate of life expectancy from the respective sources and exploit the implied annual rates of increase to construct a full and uninterrupted time series.

\bigskip

And third, to capture human capital investment, we rely on average years of schooling as a widely recognized indicator of human capital formation \citep{BarroLee2010Education}. The average years of schooling denote mean number of completed years of schooling that individuals in a given population have attained. By providing an easily tractable summary measure of the educational attainment of the population, it captures the level of education achieved over a period of time consistently. We use recently reconstructed historical estimates of the educational attainment \citep{lee2016human} to  build a concise time-series of the average years of education for our period of estimation.

\subsection{Auxilarly covariates}

To adequately capture and reproduce the outcome trajectories prior to the establishment of National Health Service (i.e. Sistema Nacional de Salud), we build a compact vector of pre-reform outcome values in most recent pre-revolution years together with the auxiliary covariates as a separate class of predictors. To this end, we consider physical geographic characteristics to capture the salience of the disease-related and climatic environment such as latitude (in absolute terms), soil quality, fraction of land area in tropical zone, fraction of land area within 300 km coastline. Second, to capture demographic salience between Cuba and the Latin American country-level donor pool, we include population density as a separate predictor based on the historical reconstruction of density estimates \citep{KleinGoldewijkEtAl2018Population} linked to the United Nations estimates of population density in the post-1950 period. Furthermore, to establish salience in terms of colonial historical background, we use recently constructed estimates of the European share of the population during colonization from \citet{easterly2016european} who demonstrate a strong association between colonial European settlement and present-day differences in economic development. And lastly, we also consider per capita GDP levels \citep{bolt2014maddison} as an additional predictor to further capture salience and similarity between Cuba and Latin American countries in terms of the level of economic development. In consequence, the full set of predictors allows us to capture both the implicit and explicit attributes of Cuba's public health and human capital formation prior to the establishment of the universal health care. In the training stage, both past values of the outcomes and the full set of auxilarly covariates explain between 70\% and 90\% of the overall variation in infant mortality, life expectancy at birth and average years of education.

\bigskip

\subsection{Notes on Measurement Errors}

Historical mortality and life expectancy data for Cuba prior to the 1950s inevitably involve measurement limitations due to incomplete coverage, heterogeneous reporting standards, and inconsistent classification of live births and infant deaths \citep{DiazBriquets1983Mortality}. To address these challenges, we combine multiple contemporaneous and retrospective sources, construct a median trajectory for each year from independent estimates \citep{alvarez1975life, farnos1977life, AstorgaFitzgerald1998}, and anchor these series to the UN Population Division's standardized post-1950 benchmarks. This approach reduces idiosyncratic noise from any single source and ensures comparability across countries in the donor pool. Remaining measurement error in early periods is likely classical rather than systematic, and thus should attenuate rather than exaggerate estimated treatment effects.

\bigskip

For infant mortality, the adoption of WHO definitions in 1965 may introduce a discrete shift in classification. However, because we align pre-1965 census-based estimates with post-1950 UN data through linked series, any discontinuity is orthogonal to treatment assignment and does not bias the synthetic control construction. Moreover, the excellent pre-intervention fit between Cuba and Synthetic Cuba across all outcomes suggests that remaining measurement noise does not meaningfully distort the validity of the identification strategy. The robustness of our findings to ASCM, SDID, IFE, and matrix completion, each of which handles measurement error differently, provides further assurance that our results are not artifacts of historical data imprecision.

\section{Identification Strategy} \label{identification}
To estimate the causal impact of Cuba's National Health Service (NHS) established in 1960 on public health outcomes and human capital formation, we employ the Synthetic Control Method (SCM) developed by \citet{abadie2003economic} and formalized in \citet{abadie2010synthetic} and \citet{abadie2015comparative}. The SCM constructs a synthetic version of Cuba by creating a weighted combination of countries from a donor pool---former European colonies in the Western Hemisphere that did not implement comprehensive universal health care during the same period. This synthetic control closely mirrors Cuba's pre-intervention characteristics and outcome trends, providing a credible counterfactual for what would have occurred in the absence of the NHS.

\bigskip

The validity of the SCM relies on the assumption that, in the absence of the intervention, the outcome trajectories of Cuba and its synthetic counterpart would have followed similar paths. By comparing the post-intervention outcomes of Cuba with those of the synthetic control, we attribute any significant deviations to the effect of the NHS.

\bigskip

To enhance the robustness of our estimates, we incorporate several complementary methods. First, we employ the Augmented Synthetic Control Method (ASCM) proposed by \citet{benmichael2021augmented}. The ASCM improves upon the traditional SCM by "augmenting" the SCM estimate with an outcome model and adding a ridge penalty to estimate the weights, therefore adjusting for any discrepancies in pre-intervention outcomes between the treated unit and the synthetic control, effectively reducing bias in the estimated treatment effect. This method is particularly useful when the fit between the treated unit and the synthetic control is imperfect during the pre-intervention period.

\bigskip

Second, we employ the Synthetic Difference-in-Differences (Synthetic DiD) approach developed by \citet{arkhangelsky2021synthetic} as an additional robustness check. The Synthetic DiD combines the strengths of the difference-in-differences estimator with the SCM. This method estimates not only unit specific weights but also give different weights to each pre-intervention period. This allows for time-varying unobserved confounders, thus  relaxing the parallel trends assumption, and non perfect fit between the treated and its synthetic counterpart therefore overcoming limitation of the SCM.

Third, we implement the interactive fixed effects (IFE) estimator developed by \citet{xu2017generalized}, which models the counterfactual using a factor model framework. Unlike SCM, which relies on a weighted average of control units, IFE allows for time-varying unobserved heterogeneity by estimating latent factors that capture common trends across units. This approach is particularly valuable when the parallel trends assumption may be violated, as it can accommodate unit-specific responses to unobserved common shocks.

\bigskip

Finally, we apply the matrix completion method proposed by \citet{athey2021matrix}, which treats the missing counterfactual outcomes as entries in a partially observed matrix. This approach leverages regularized principal component analysis to impute the missing values, exploiting both cross-sectional and time-series patterns in the data. By using nuclear norm regularization with varying penalty parameters ($\lambda$), the method provides a flexible framework for estimating treatment effects that is robust to model misspecification and can handle complex patterns in the data structure.

\bigskip

\bigskip

By utilizing these complementary methodologies, we ensure that our findings are robust to various identification strategies and that the estimated effects are credibly attributed to the introduction of the NHS in Cuba. These methods collectively address potential confounding factors and strengthen the causal interpretation of our results.

\subsection{Threats to identification and how we address them}\label{sec:threats}

Three identification concerns are worth discussing explicitly: the bundled nature of the post-1959 reforms, post-revolutionary mass emigration, and post-1990 macroeconomic shocks.

\paragraph{Bundled reforms.} As discussed in Section~\ref{history}, the 1961 establishment of the NHS coincided with the National Literacy Campaign and was embedded in a broader bundle of revolutionary changes (land reform, nationalization, the U.S.\ embargo, and Soviet support). Our estimates therefore identify the joint effect of universal health-care provision and the contemporaneous mass-literacy reform, rather than that of either component in isolation. We adopt this bundled interpretation throughout. The literacy campaign is, however, a one-shot intervention that lifted the literacy rate from $\sim$60--80 percent to $\sim$96 percent within eight months; its mechanical contribution to subsequent years-of-schooling is bounded by that one-time level shift and cannot mechanically generate a divergence that continues to widen for six decades. The persistent, cumulative pattern we document for schooling is therefore more consistent with the joint operation of permanently expanded health-care access and an upgraded education system than with the literacy push alone.

\paragraph{Mass emigration.} Between 1959 and 2010, an estimated 1.4--1.6 million Cubans emigrated \citep{duany2017cuban,ajadiaz2014cuban}, with the largest waves occurring in 1959--62 (largely upper-class and professional households), 1965--73 (\textit{Vuelos de la Libertad}), 1980 (Mariel boatlift), and the post-1994 \textit{balseros}. Because the emigrants were on average wealthier, better educated, and lighter-skinned than the residual population, selection in principle works in two directions for our outcomes. On the one hand, the residual population is more rural, lower-income, and less educated, biasing comparisons of infant mortality and schooling \textit{against} finding gains, since these are precisely the groups for which baseline outcomes were worst. On the other hand, the loss of high-mortality adult cohorts could mechanically lower observed mortality. Quantitatively, however, the share of cumulative emigration relative to the total Cuban population is small in the years where our infant-mortality and schooling gaps emerge (roughly 5--10 percent through 1980), and the demographic composition of the residual population continued to age in line with Latin American comparators. We view it as unlikely that emigration of an even more healthy-than-average emigrant pool could account for the observed 15--29 percent drop in infant mortality or a 1.5--2 year rise in schooling relative to synthetic Cuba. Even under a conservative assumption that emigrants would have had infant mortality rates 50 percent below the population average, removing 10 percent of the population at random raises the mechanical estimate of the post-treatment gap by only a few percent of its observed magnitude. We therefore interpret emigration as a level-shifting nuisance that reinforces rather than overturns our central findings.

\paragraph{Post-1990 macroeconomic shocks.} The 1989--91 collapse of the Soviet Union triggered the so-called Special Period in Cuba: GDP per capita fell by roughly one-third, food and medicine imports collapsed, and adult caloric intake dropped substantially \citep{garfield1997impact, barry2000effect}. These shocks plausibly attenuate any positive long-run effect of the 1961 reforms, particularly for outcomes such as adult life expectancy that respond to caloric intake, chronic-disease management, and pharmaceutical availability. In contrast, infant mortality is shaped more directly by maternal care, vaccination coverage, sanitation, and primary-care infrastructure, all of which were defended as state priorities during the Special Period. The differential pattern we observe---a stable infant-mortality and schooling gap alongside a post-1990 attenuation in the life-expectancy gap---is therefore consistent with the Soviet-collapse channel rather than with a generic decay of the reform's effects. Our matrix-completion, IFE, and ASCM specifications also generate noticeably more uncertainty for life expectancy than for infant mortality and schooling, a pattern that aligns with this interpretation. A direct test of the Soviet-collapse interpretation is to split the post-treatment window at 1990 and re-estimate the SCM gap separately for 1961--1990 (pre--Special Period) and 1990--2022 (post--Special Period). Under our interpretation, the pre-1990 life-expectancy gap should be large and statistically comparable in relative magnitude to the infant-mortality gap, while the post-1990 gap should attenuate sharply for life expectancy but not for infant mortality or schooling. We report this split as a complementary robustness check in the results section and, where data permit, augment it with a difference-in-differences-style placebo test in which we artificially declare 1990 as the treatment date for the donor pool to confirm that the post-1990 contraction in Cuban life expectancy reflects a Cuba-specific shock rather than a region-wide trend.

\section{Results}\label{sec:results}
\subsection{Construction of Synthetic Cuba and Pre-Intervention Fit}
Table \ref{tab:synthetic_control} reports the composition of the synthetic control for each outcome and the corresponding balance on pre-intervention predictors. The synthetic Cuba for infant mortality is a convex combination of El Salvador (62\%), the United States (27\%), and Venezuela (11\%). This combination captures the fact that, prior to the early 1960s, Cuba's mortality profile lay between that of relatively poor Latin American countries and that of a richer, high-capacity health system. For life expectancy, the optimal synthetic control group is primarily composed of Paraguay (69\%) and Uruguay (30\%), two countries that jointly replicate Cuba's intermediate position between upper-middle-income Latin America and the Southern Cone. The synthetic trajectory for average years of schooling is constructed from Costa Rica (44\%), Panama (37\%), and the Dominican Republic (19\%), mirroring Cuba's comparatively high but not exceptional educational attainment in the pre-reform period.

\bigskip

The quality of the pre-intervention fit is central to our identification strategy. For each outcome, we include multiple pre-1961 lags of the dependent variable as well as a rich set of time-invariant predictors: absolute latitude, population density, the share of European descent, GDP per capita (or its logarithm), distance to the coast, soil quality, and the fraction of land in the tropical zone. As shown in Table \ref{tab:synthetic_control} , the synthetic controls match Cuba very closely on these dimensions and substantially improve on the donor pool averages. For instance, the lagged values of infant mortality for Cuba and synthetic Cuba are nearly indistinguishable over the four pre-treatment periods reported, while both differ markedly from the average donor-pool country. A similarly tight alignment is evident for life expectancy and schooling. The pre-intervention levels and trends for synthetic Cuba track the actual series closely, with discrepancies that are small in magnitude relative to the cross-sectional dispersion in the donor pool. This pattern indicates that the synthetic controls reproduce not only the level but also the slope of Cuba's pre-reform trajectories, which is critical for credible extrapolation into the post-reform period.

{\scriptsize
\begin{longtable}{@{\extracolsep{\fill}}llccc@{}}
\caption{Synthetic Control Weights and Balance by Outcome} \label{tab:synthetic_control} \\
\toprule
\textbf{Outcome} & \textbf{Variable/Country} & \textbf{Cuba} & \textbf{Synthetic Cuba} & \textbf{Donor Sample / Weight} \\
\midrule
\endfirsthead

\endhead

\endfoot

\bottomrule
\endlastfoot

\textbf{Infant Mortality} & \multicolumn{4}{l}{\textbf{Weights}} \\
& El Salvador & & & 0.62 \\
& United States of America & & & 0.27 \\
& Venezuela & & & 0.11 \\
\cmidrule(l){2-5}
& \multicolumn{4}{l}{\textbf{Balance}} \\
& Infant Mortality (t-1) & 4.78 & 5.13 & 5.14 \\
& Infant Mortality (t-2) & 4.82 & 5.14 & 5.16 \\
& Infant Mortality (t-3) & 4.85 & 5.16 & 5.18 \\
& Infant Mortality (t-4) & 4.88 & 5.17 & 5.20 \\
& Absolute Latitude & 21.63 & 21.62 & 22.92 \\
& Population Density & 0.04 & 0.05 & 0.04 \\
& \% European Descent & 63.00 & 57.46 & 53.40 \\
& GDP per capita (log) & 7.72 & 8.49 & 8.46 \\
& Infant Mortality & 5.73 & 5.71 & 5.66 \\
& Longitude & -79.02 & -92.87 & -83.79 \\
& Near Coast (Dummy) & 100.00 & 68.48 & 40.35 \\
& Soil Quality Index & 45.95 & 54.61 & 38.42 \\
& \% Tropical Area & 100.00 & 72.82 & 49.15 \\
\midrule
\textbf{Life Expectancy} & \multicolumn{4}{l}{\textbf{Weights}} \\
& Paraguay & & & 0.69 \\
& Uruguay & & & 0.30 \\
\cmidrule(l){2-5}
& \multicolumn{4}{l}{\textbf{Balance}} \\
& Life Expectancy (t-1) & 62.08 & 61.22 & 56.36 \\
& Life Expectancy (t-2) & 60.11 & 60.50 & 53.86 \\
& Life Expectancy (t-4) & 45.00 & 46.71 & 44.97 \\
& Absolute Latitude & 21.63 & 25.86 & 22.50 \\
& Population Density & 0.04 & 0.00 & 0.01 \\
& \% European Descent & 63.00 & 64.45 & 60.14 \\
& GDP per capita (log) & 7.73 & 8.01 & 8.16 \\
& Life Expectancy & 46.06 & 45.92 & 43.75 \\
& Longitude & -79.02 & -57.88 & -73.40 \\
& Near Coast (Dummy) & 100.00 & 11.58 & 29.28 \\
& Soil Quality Index & 45.95 & 40.71 & 34.89 \\
& \% Tropical Area & 100.00 & 40.50 & 47.21 \\
\midrule
\textbf{Education} & \multicolumn{4}{l}{\textbf{Weights}} \\
& Costa Rica & & & 0.44 \\
& Panama & & & 0.37 \\
& Dominican Republic & & & 0.19 \\
\cmidrule(l){2-5}
& \multicolumn{4}{l}{\textbf{Balance}} \\
& School Enrollment (t-1) & 4.08 & 4.04 & 3.82 \\
& School Enrollment (t-2) & 4.00 & 3.98 & 3.76 \\
& School Enrollment (t-3) & 3.91 & 3.93 & 3.70 \\
& School Enrollment (t-4) & 3.83 & 3.87 & 3.64 \\
& Absolute Latitude & 21.63 & 11.33 & 20.36 \\
& Population Density & 0.04 & 0.02 & 0.02 \\
& \% European Descent & 63.00 & 53.35 & 52.68 \\
& GDP per capita (log) & 7.55 & 7.67 & 7.79 \\
& Longitude & -79.02 & -80.15 & -77.13 \\
& Near Coast (Dummy) & 100.00 & 99.57 & 45.60 \\
& Population in 1400 & 95479.00 & 65576.68 & 627850.25 \\
& School Enrollment & 1.90 & 1.91 & 1.93 \\
& Soil Quality Index & 45.95 & 46.16 & 37.31 \\
& \% Tropical Area & 100.00 & 94.66 & 61.54 \\
\end{longtable}
}

Some differences remain on a subset of predictors, most notably in GDP per capita for the infant-mortality specification and in certain geographical characteristics such as longitude, coastal exposure, and the share of tropical land. These discrepancies are modest in economic terms and go in both directions: for some variables, synthetic Cuba slightly exceeds the Cuban value, while for others it falls below. Importantly, they are small relative to the variation across donor countries and do not translate into visible misfit in the pre-intervention outcome paths, where all three outcomes exhibit a close tracking between Cuba and its synthetic counterpart. Nonetheless, we treat these residual imbalances conservatively. In the sections that follow, we complement the baseline synthetic-control estimates with a battery of alternative estimators, augmented synthetic control, synthetic difference-in-differences, interactive fixed effects, and matrix completion. This particular suite of estimator is explicitly designed to adjust for remaining discrepancies in pre-intervention fit. The convergence of results across these methods provides an additional layer of reassurance that our findings are not an artifact of imperfect balance in the baseline specification.

\subsection{Main Results}

Table \ref{tab:synthetic_control_results} summarizes the baseline treatment effects obtained from the synthetic control method for our three main outcomes: infant mortality, life expectancy at birth, and average years of education. The estimates display a clear pattern. First, the establishment of universal health care and the associated reforms is followed by a sizeable and persistent decline in infant mortality relative to synthetic Cuba. Second, life expectancy improves in levels and attains a higher trajectory than its synthetic counterpart, but the effect is somewhat more modest in percentage terms and attenuates over time. Third, average years of education exhibit the largest proportional response: a sustained and widening divergence from synthetic Cuba that is consistent with a structural shift in human capital accumulation.

\bigskip

On average over the post-treatment period, infant mortality in Cuba is 0.84 deaths per 1,000 live births lower than in synthetic Cuba, corresponding to a 15\% reduction relative to the synthetic counterfactual. Life expectancy at birth is on average 2.93 years higher, a gain of roughly 6\%, while average years of education are 1.84 years higher, implying a 96\% increase relative to the synthetic benchmark. In terms of dynamics, the infant mortality gap widens throughout the first two post-reform decades and then stabilizes at a lower level, the life expectancy gap peaks around 1980 before partially narrowing, and the education gap grows monotonically and remains large even in the most recent decade. These patterns align closely with the mechanisms discussed in Section~\ref{sec:theory}, displaying early and persistent gains in survival, a substantial but shock-sensitive improvement in longevity, and a gradual yet durable reorientation of the human capital trajectory.

\begin{longtable}{lccc}
\caption{Summary of Treatment Effects} \label{tab:synthetic_control_results} \\
\toprule
 & \multicolumn{3}{c}{Outcome Variable} \\
 \cmidrule{2-4}
 & Infant Mortality & Life Expectancy & Years of Education \\
 & (1) & (2) & (3) \\
\midrule
\endfirsthead

\endhead

\endfoot

\bottomrule
\endlastfoot

Average Effect (Post-Treatment) & -0.84 & +2.93 & +1.84 \\
 & (-15\%) & (+6\%) & (+96\%) \\
Effect vs. Last Pre-Treatment & +17\% & +4.7\% & +45\% \\
\midrule
\multicolumn{4}{l}{\textit{Dynamic Effects by Decade:}} \\
1970 & -0.583 & +3.85 & +0.92 \\
1980 & -1.096 & +6.22 & +1.56 \\
1990 & -1.038 & +4.50 & +1.84 \\
2000 & -0.923 & +4.82 & +1.62 \\
2010 & -0.863 & +4.26 & +2.20 \\
2020 & -0.827 & +2.83 & +2.01 \\
\end{longtable}

\subsubsection{Infant Mortality}

\begin{figure}[H]
\centering
\includegraphics[width=0.9\linewidth]{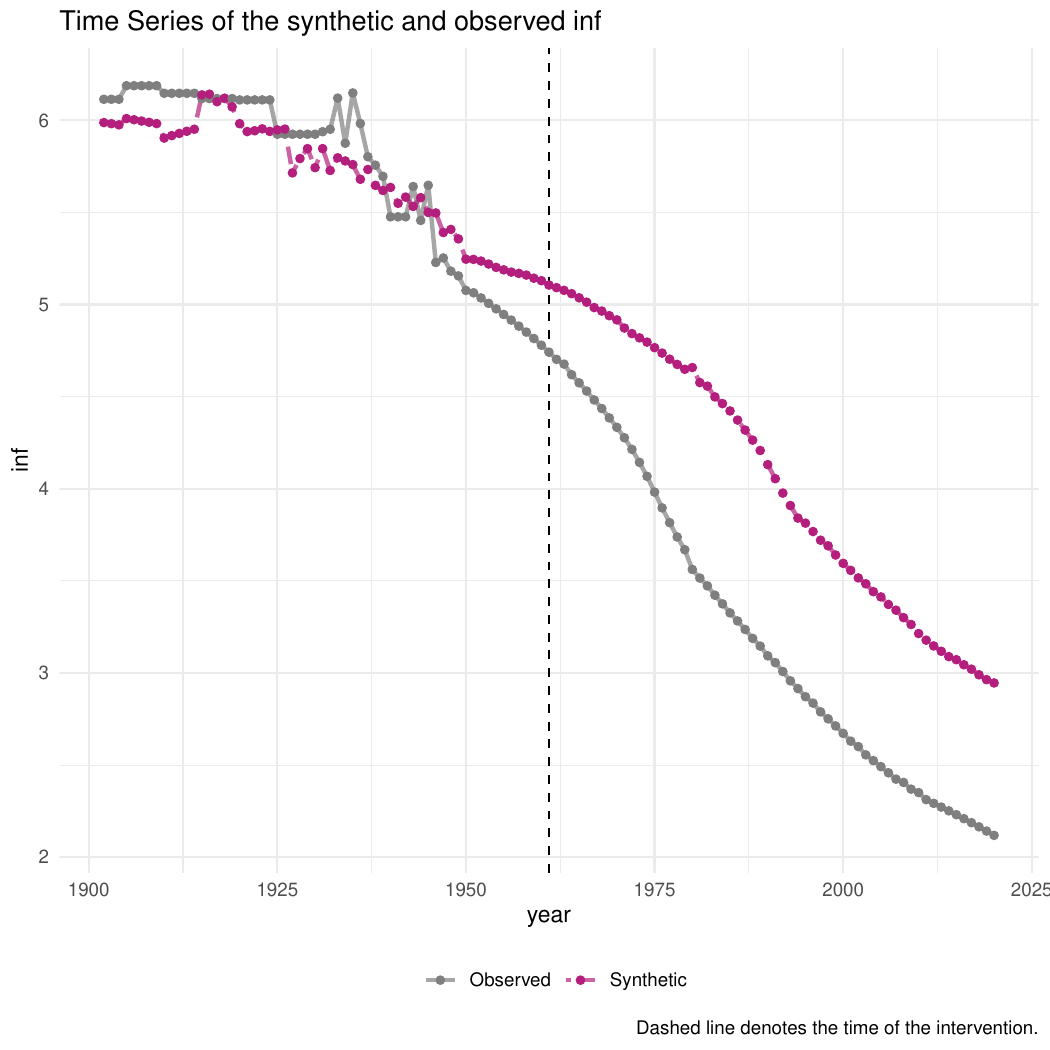}
\caption{Cuba vs Synthetic Cuba Infant Mortality Trajectories. }
\label{fig:scminf}
\end{figure}

Figure \ref{fig:scminf} plots the time paths of Cuba's infant mortality rate and that of synthetic Cuba from 1900 to 2022. The pre-intervention fit is extremely tight: from the early twentieth century through 1960, the synthetic control tracks the Cuban series almost exactly, reflecting the close matching of both levels and trends achieved by the weighting procedure. The intervention year is indicated by a vertical dashed line. Immediately after the introduction of the health reform, the observed Cuban trajectory begins to diverge downward from the synthetic counterfactual.

\bigskip

The magnitude of the divergence is economically meaningful. As reported in Table~\ref{tab:synthetic_control_results}, the average post-treatment gap amounts to 0.84 infant deaths per 1,000 live births, or roughly a 15\% reduction relative to synthetic Cuba. Measured against the last pre-treatment level, this corresponds to a 17\% improvement. The gap does not arise from a transient deviation. It widens during the first twenty years after the reform---reaching 0.58 deaths per 1,000 by 1970 and 1.10 by 1980, and then stabilizes in the range of 0.83-0.93 deaths per 1,000 from 2000 onward. The synthetic trajectory itself continues to decline over time, reflecting broader improvements in health across the donor pool, but Cuba's decline is steeper and sustained. Taken together, these patterns indicate that the health reform is associated with a substantial and persistent reduction in infant mortality beyond what would be expected given regional trends and the pre-existing trajectory of Cuban health outcomes. The timing, magnitude, and durability of the gap are consistent with the view that universal access to preventive and primary care, combined with enhanced state capacity and improved early-life conditions, permanently shifted the survival environment for newborns and infants.

\subsubsection{Life Expectancy}

\begin{figure}[H]
    \centering
    \includegraphics[width=0.9\linewidth]{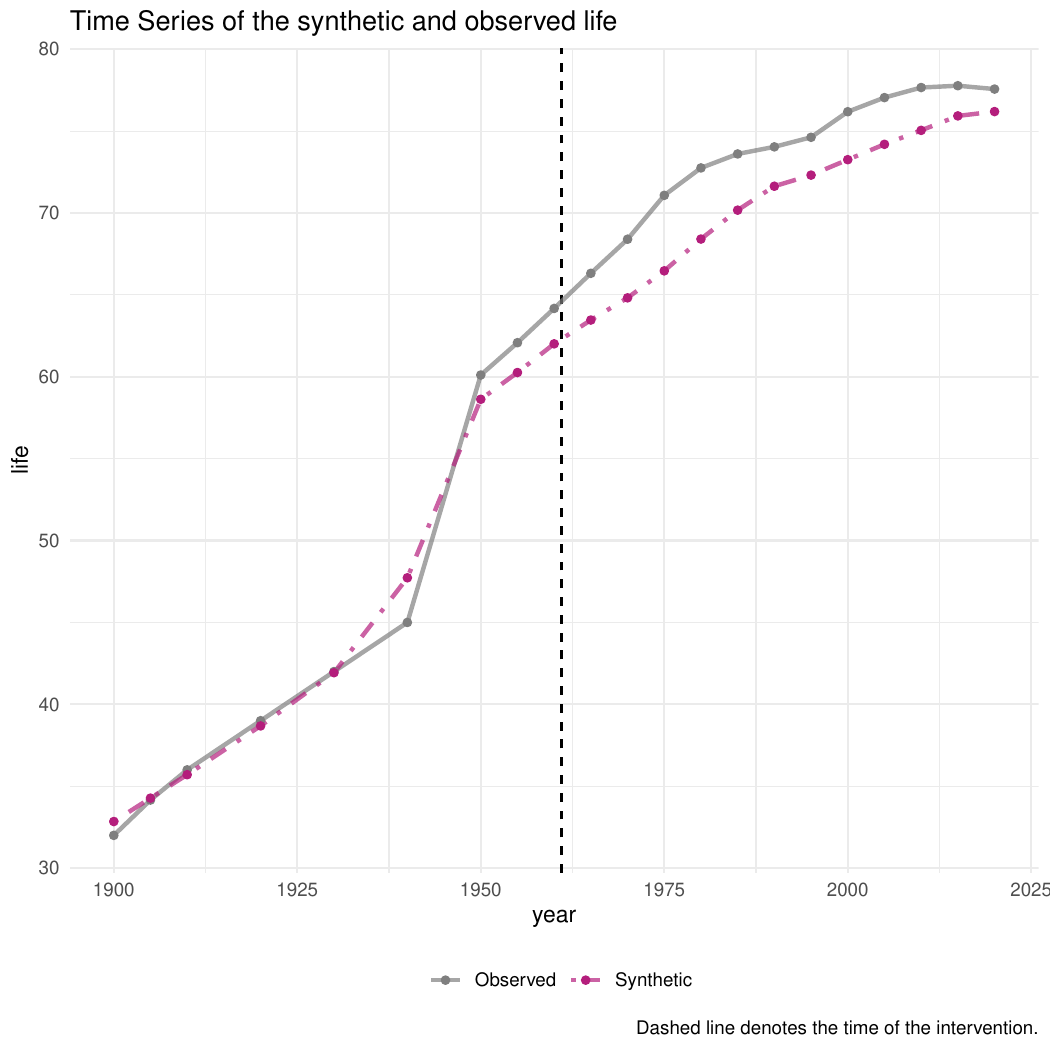}
    \caption{Cuba vs Synthetic Cuba Life Expectancy Trajectories. }
    \label{fig:scmlife}
\end{figure}

Figure \ref{fig:scmlife} presents the corresponding actual and synthetic trajectories for life expectancy at birth. As with infant mortality, the synthetic control reproduces the Cuban pre-intervention path very closely. Both series move almost in lockstep from 1900 to 1960, confirming that the donor-weighting scheme captures the level and slope of Cuba's pre-reform longevity profile. Following the introduction of universal health care, observed life expectancy in Cuba rises above that of synthetic Cuba, with the gap reaching its maximum in the decades immediately after the reform.

\bigskip

The average post-treatment effect reported in Table~\ref{tab:synthetic_control_results} is an increase of 2.93 years, or about 6\% relative to the synthetic counterfactual. The effect is not constant over time. It grows from 3.85 years in 1970 to 6.22 years by 1980, before declining to 4.50 years in 1990 and to 2.83 years by 2020. This pattern suggests that the reform shifted the level of life expectancy upward, but that subsequent macroeconomic and geopolitical shocks, most notably the collapse of the Soviet Union and the associated `Special Period'', moderated the long-run divergence in longevity. The attenuation is consistent with the idea that life expectancy, unlike infant mortality, is influenced not only by the quality and reach of the health system but also by broad economic conditions, resource constraints, and changes in adult mortality risks. Thus, the life expectancy results point to a substantial but more nuanced long-run impact. Universal health care appears to have moved Cuba onto a higher longevity path relative to comparable countries, but the cumulative effect is smaller in relative terms than for infant mortality and education, and more sensitive to later shocks. This heterogeneity across outcomes is precisely what the theoretical framework in Section~\ref{sec:theory} would lead one to expect.

\subsubsection{Education}
\begin{figure} [H]
    \centering
    \includegraphics[width=0.9\linewidth]{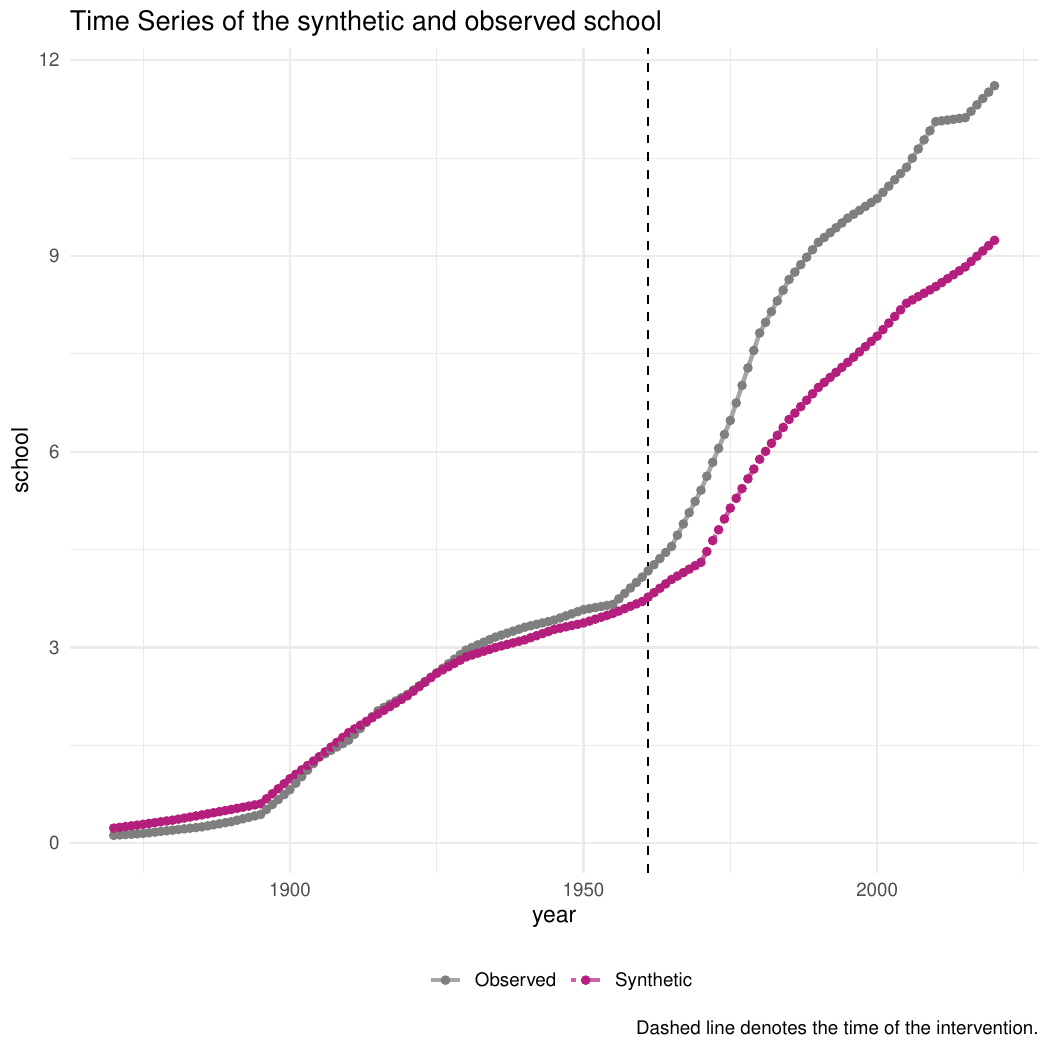}
    \caption{Cuba vs Synthetic Cuba Average Years of Education Trajectories. }
    \label{fig:scmschool}
\end{figure}

Figure \ref{fig:scmschool} displays the trajectories for average years of education. As with the health outcomes, the synthetic control delivers an excellent pre-intervention fit. The Cuban and synthetic series are nearly indistinguishable up to 1960, indicating that the donor combination of Costa Rica, Panama, and the Dominican Republic accurately reproduces both the level and trend of Cuban schooling in the pre-reform era. After the introduction of universal health care and the mass literacy and education reforms, the Cuban schooling trajectory diverges sharply from its synthetic counterpart. According to Table~\ref{tab:synthetic_control_results}, the average effect over the post-treatment period is an increase of 1.84 years of schooling, corresponding to a 96\%  gain relative to synthetic Cuba. When compared to the last pre-treatment value, this implies a 45\% increase. The gap grows steadily over time, namely, from 0.92 years in 1970 to 1.56 years in 1980, 1.84 years in 1990, and a peak of 2.20 years by 2010, remaining at 2.01 years in 2020. Unlike life expectancy, there is no evidence of a reversal or convergence in the late period. Instead, the schooling differential appears to be both large and persistent.
\bigskip
These dynamics are consistent with a structural change in the process of human capital accumulation. Improvements in early life health and survival raise the returns to schooling, the literacy campaign removes basic literacy constraints among parents and older cohorts, and the expansion of educational infrastructure enables successive cohorts to complete more years of schooling. The synthetic-control estimates suggest that this joint health-education state-building project shifted Cuba onto a permanently higher schooling path than the one it would have followed in the absence of the reform.

\bigskip

In sum, the main results from the synthetic control analysis reveal a coherent pattern of a large and enduring reduction in infant mortality, a substantial but shock-sensitive increase in life expectancy, and a dramatic, persistent rise in educational attainment. In the following sections, we show that these findings are robust to a wide range of alternative estimators and placebo exercises, and we provide inference procedures that quantify the statistical significance of the estimated gaps.

\subsection{Inference}

The synthetic control method is designed for settings with a small number of treated units, often only one, and a modestly sized donor pool. In such environments, conventional large-sample inference is inappropriate, and the distribution of estimators cannot be reliably approximated by standard asymptotic arguments. Following \citet{abadie2010synthetic, abadie2015comparative}, we therefore base inference on placebo tests in space, in which the treatment is sequentially reassigned to each country in the donor pool, and on the comparison of post/pre-intervention mean squared prediction error (MSPE) ratios between Cuba and the donor units. These procedures exploit the finite-sample variation in the data to assess how unusual Cuba's estimated treatment effects are, relative to the distribution of `effects'' one would obtain under the null of no effect.
\bigskip
Our inferential strategy proceeds in three steps. First, for each outcome, we construct a synthetic control for every country in the donor pool as if that country had experienced the Cuban reform in 1961. For each placebo unit, we compute the gap between observed outcomes and its synthetic control over time. Second, we compare the Cuban gap to the distribution of placebo gaps, paying particular attention to their relative magnitudes in the post-intervention period. Third, to account for heterogeneity in pre-intervention fit across countries, we follow \citet{abadie2010synthetic} and compute, for each unit, the ratio of post- to pre-intervention MSPE. The MSPE ratio captures the extent to which the fit deteriorates after the intervention date. Large MSPE ratios for Cuba, relative to the donor units, indicate that the divergence observed in Cuba is unlikely to be driven by noise alone.

\begin{figure}[H]
\centering
\begin{subfigure}[b]{0.32\textwidth}
\centering
\includegraphics[width=\textwidth]{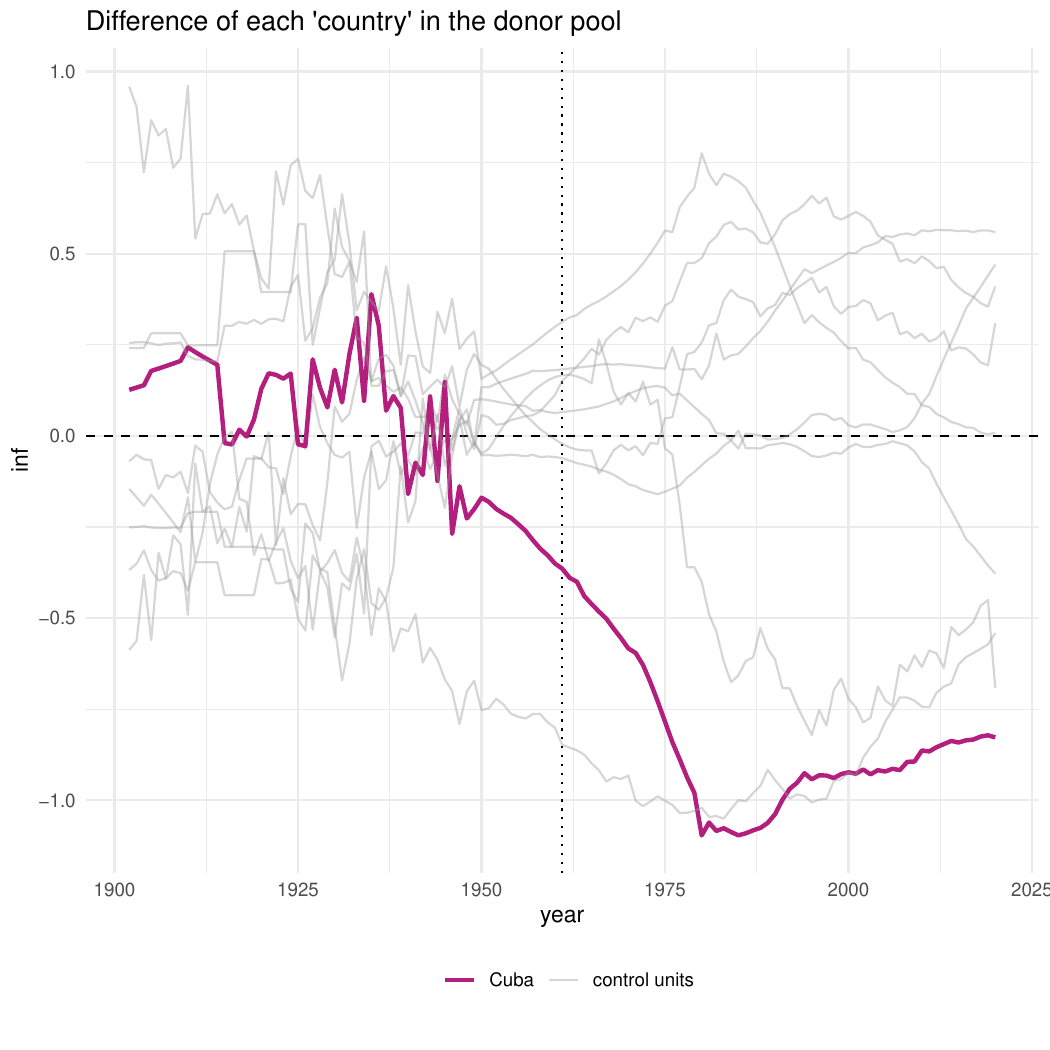}
\caption{Infant Mortality}
\label{fig:scminf_placebo}
\end{subfigure}
\hfill
\begin{subfigure}[b]{0.32\textwidth}
\centering
\includegraphics[width=\textwidth]{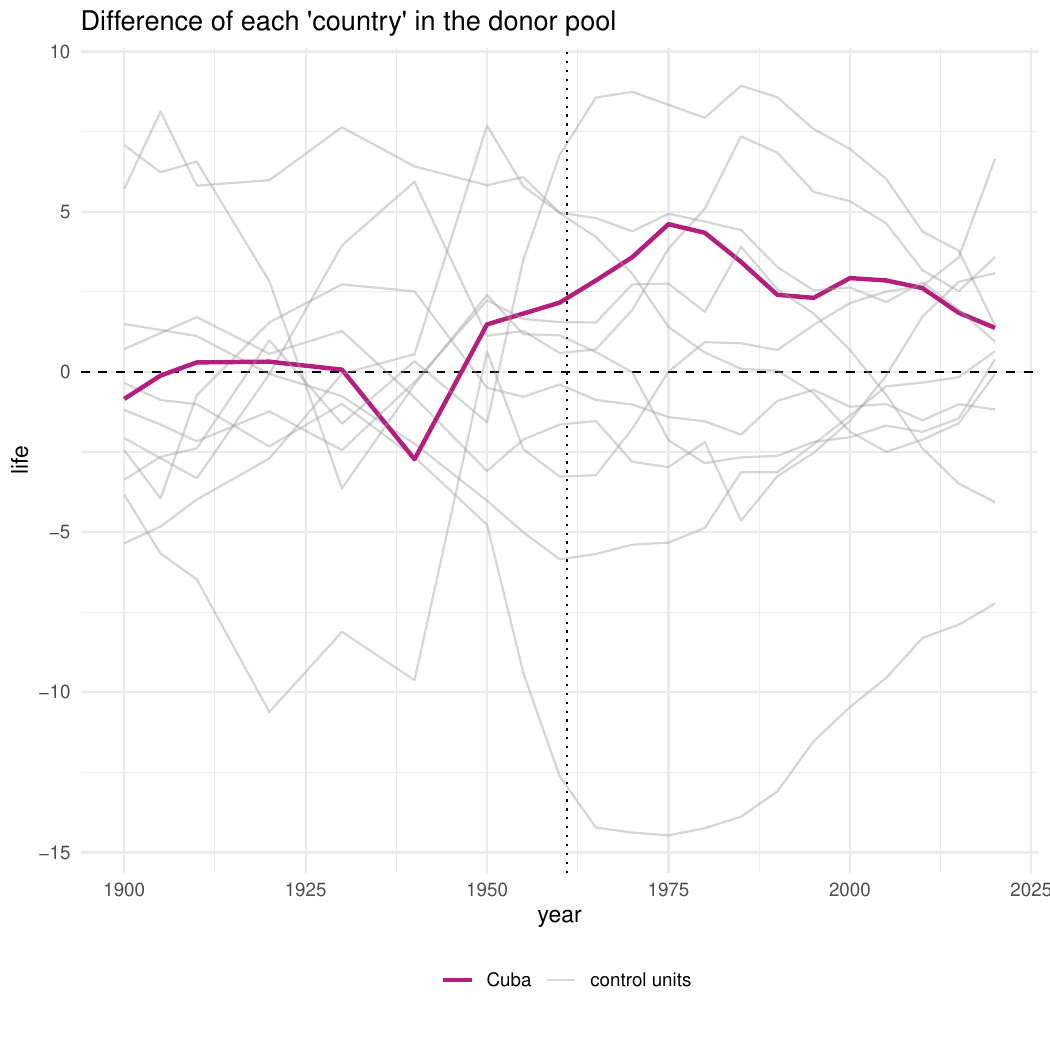}
\caption{Life Expectancy}
\label{fig:scmlife_placebo}
\end{subfigure}
\hfill
\begin{subfigure}[b]{0.32\textwidth}
\centering
\includegraphics[width=\textwidth]{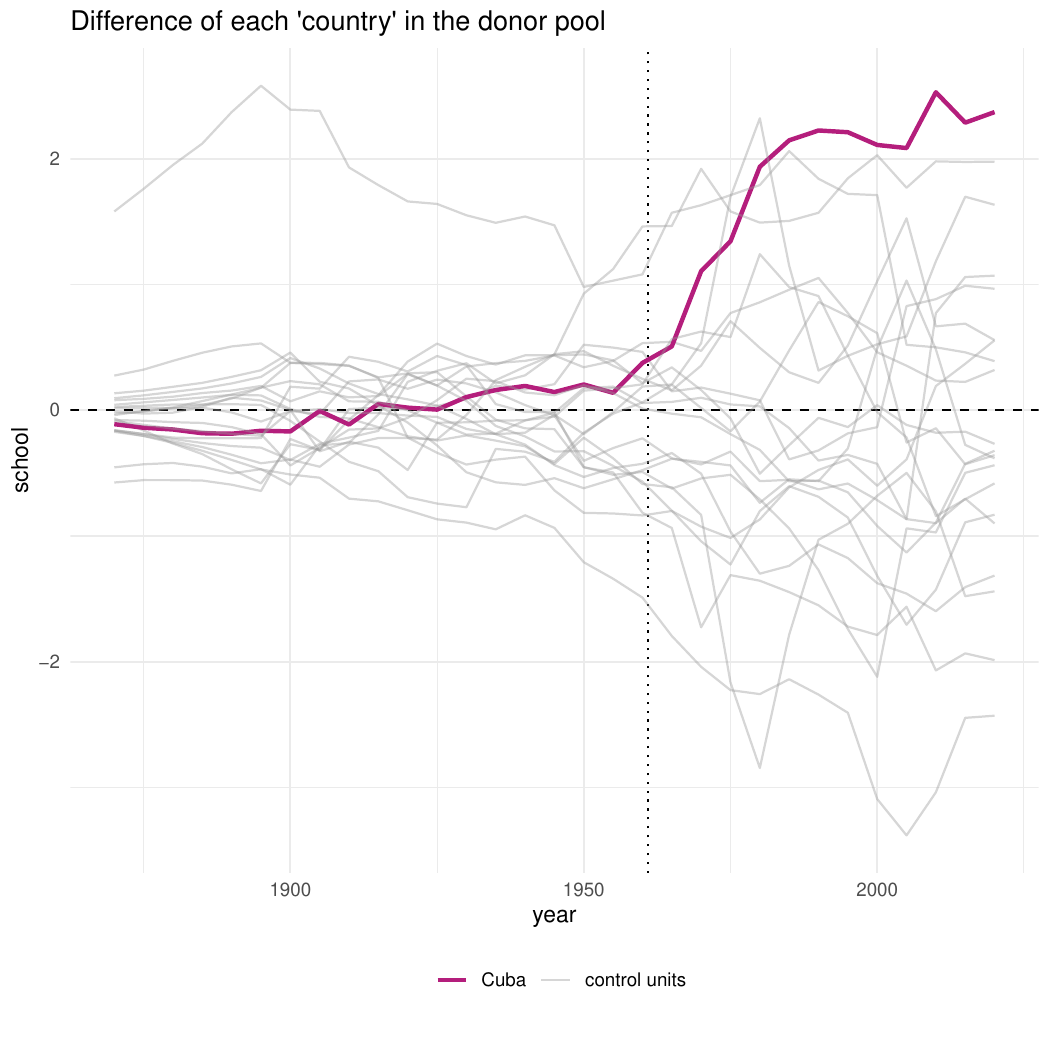}
\caption{Education}
\label{fig:scmschool_placebo}
\end{subfigure}
\caption{Placebo Tests in Space}
\label{fig:placebo_all}
\end{figure}
\begin{figure}[H]
\centering
\begin{subfigure}[b]{0.32\textwidth}
\includegraphics[width=\textwidth]{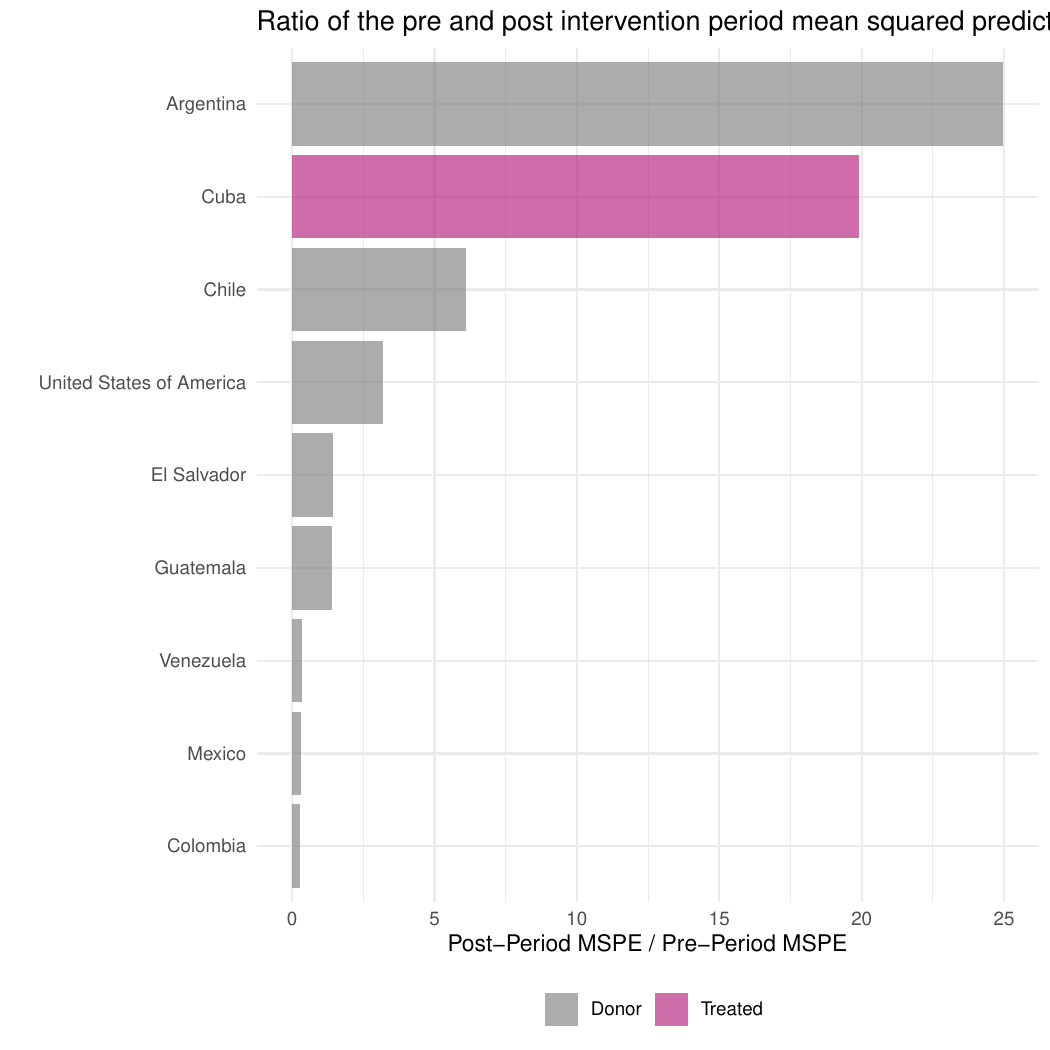}
\caption{Infant Mortality}
\end{subfigure}
\begin{subfigure}[b]{0.32\textwidth}
\includegraphics[width=\textwidth]{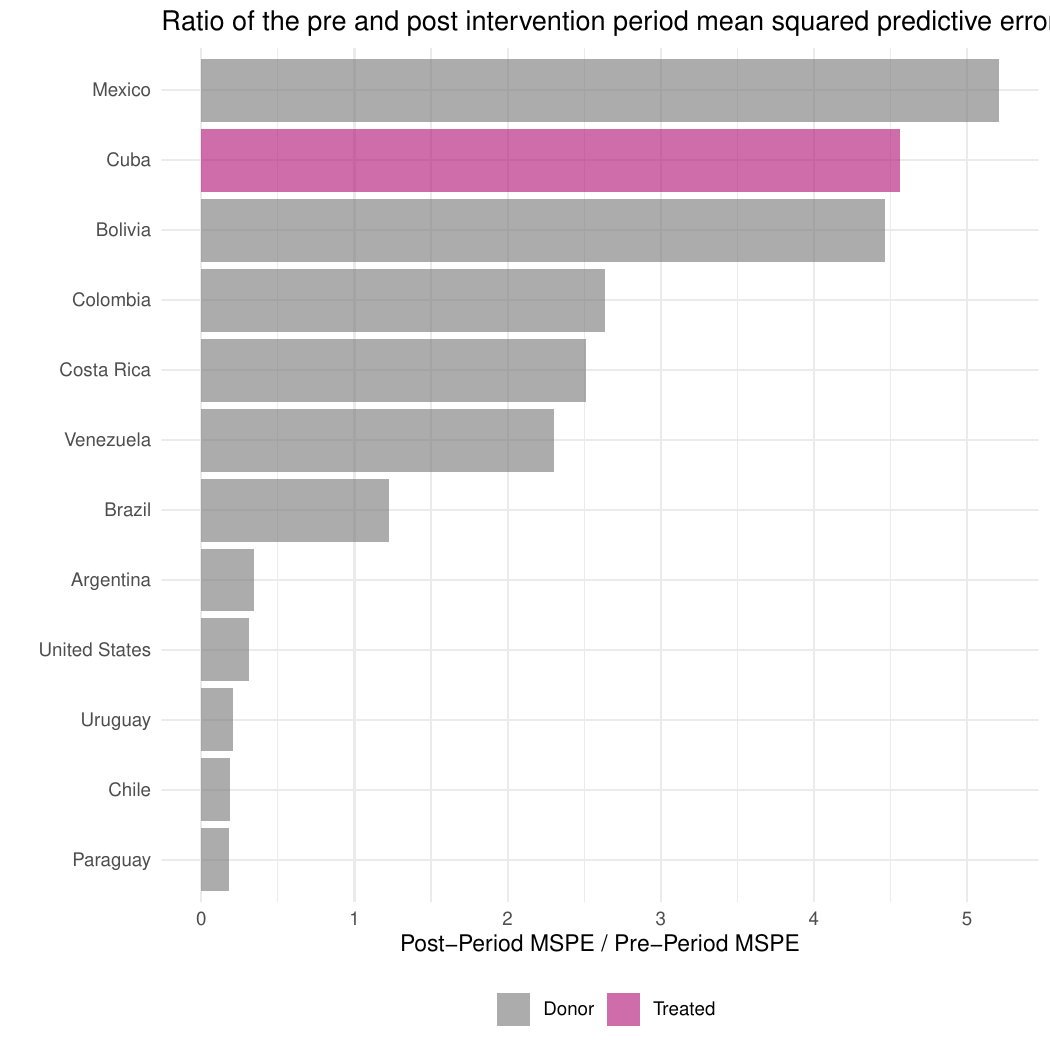}
\caption{Life Expectancy}
\end{subfigure}
\begin{subfigure}[b]{0.32\textwidth}
\includegraphics[width=\textwidth]{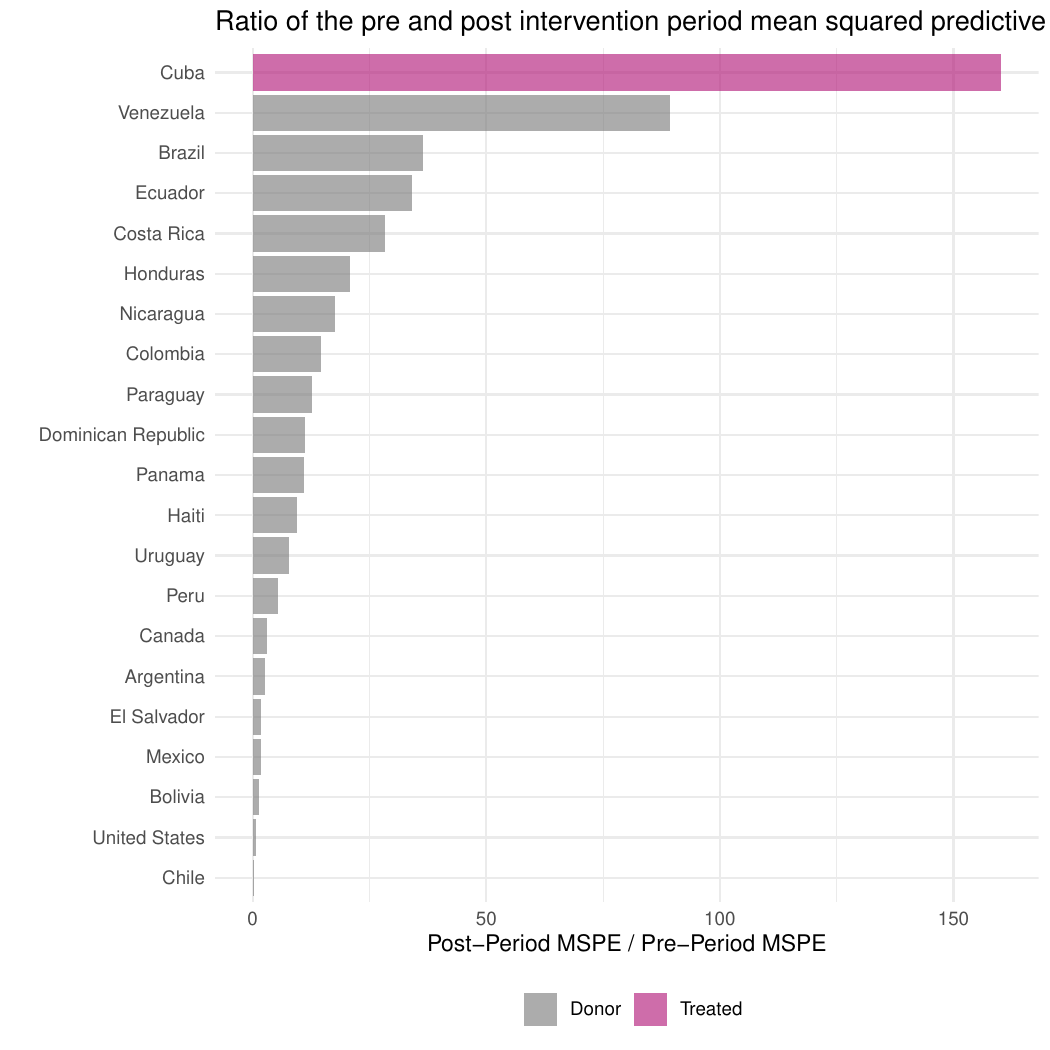}
\caption{Education}
\end{subfigure}
\caption{MSPE Ratios: Post/Pre-Intervention Period}
\label{fig:mspe}
\end{figure}

Figure \ref{fig:placebo_all} reports the placebo trajectories for all donor countries together with the Cuban gap for each outcome  and Figure \ref{fig:mspe} reports the MSPE ratios for Cuba and all donor countries for each outcome.  For each unit---Cuba and all donors---we plot the difference between the observed outcome and its synthetic control over the entire 1900-2022 period, with the intervention date marked by a vertical line. Under the null hypothesis that the reform had no effect, Cuba's post-1961 gap should not be systematically larger in magnitude than the placebo gaps for donor countries that did not implement comparable universal health-care reforms.

\bigskip

For infant mortality (panel 4a), Cuba's actual trajectory stands out. Before 1961, the gaps for all countries---including Cuba---oscillate around zero with small amplitude, reflecting good pre-treatment fit. After 1961, Cuba's gap becomes distinctly negative and remains so for the remainder of the sample period, whereas placebo gaps for donor countries display no systematic, persistent divergence of similar magnitude. Thus, the evidence is consistent exactly what one would expect if the reform produced a sustained reduction in infant mortality. The actual Cuban line separates from the `cloud'' of placebo lines in the post-intervention period while remaining indistinguishable from them beforehand.

\bigskip

For life expectancy (panel 4b), the Cuban line again drops out of the pre-1961 bundle and moves above most placebo trajectories after the reform, indicating an improvement relative to synthetic control. However, the separation is less pronounced than for infant mortality and exhibits more overlap with placebo paths in later decades. This visual impression foreshadows the more nuanced inference we obtain for life expectancy where we find evidence of a positive effect, but it is weaker and more sensitive to later shocks than for infant mortality. For education (panel 4c), the Cuban trajectory once more exhibits a clear post-1961 divergence, this time upward. The gaps for donor countries remain close to zero or fluctuate without clear trend, while the Cuban gap grows steadily and remains positive throughout the post-reform period. The contrast between Cuba and the placebo units is especially stark in the later years, consistent with the interpretation that the health and education reform set Cuba on a permanently higher schooling path.

\bigskip

These placebo plots implement a finite-sample version of a randomization test. We compare the `treatment effect'' for the actually treated unit (Cuba) to the distribution of effects obtained when the treatment is reassigned to donor units that were not exposed to the policy. If the Cuban effect were typical of this distribution, we would regard it as compatible with chance. The fact that Cuba's post-intervention gap trajectory is consistently among the most extreme lines in panels (a) and (c) suggests that the observed declines in infant mortality and increases in schooling are unlikely to be purely spurious.

\bigskip

A key concern in interpreting placebo tests is that some donor countries may exhibit poor pre-treatment fit. More specifically, if a synthetic control cannot reproduce Cuba's pre-1961 trajectory, then large post-1961 gaps for that country provide little information about the plausibility of the treatment effect under the null hypothesis. To address this, \citet{abadie2010synthetic, abadie2015comparative} propose focusing on the ratio of post- to pre-intervention MSPE. This ratio is small when the synthetic control tracks the treated unit well both before and after the intervention and becomes large when the post-intervention discrepancies exceed what would be expected based on pre-intervention fit.

\bigskip

Against this backdrop, Figure~\ref{fig:mspe} reports the MSPE ratios for Cuba and all donor countries for each outcome. For infant mortality (\textit{panel 5a}), Cuba's MSPE ratio lies at the upper end of the donor distribution. Only a small number of placebo units exhibit comparable or larger ratios which precisely indicates that the synthetic control's fit deteriorates much more sharply for Cuba after 1961 than for the vast majority of donor countries, consistent with a genuine treatment effect. Interpreting the MSPE ratios in the spirit of the permutation test, the Cuban ratio for infant mortality corresponds to a permutation-style p-value below 0.10, meaning that fewer than 10 percent of donor units have post-/pre-MSPE ratios as large as Cuba's.

\bigskip

For education (\textit{panel 5c}), the Cuban MSPE ratio is even more extreme. The post-reform divergence in schooling generates a sharp increase in MSPE relative to the pre-reform period, while most donor countries display ratios close to one. That is, their fit does not deteriorate markedly after the artificial `treatment'' date. The Cuban observation again falls in the upper tail of the donor distribution, implying a permutation-style p-value comfortably below conventional thresholds (roughly p < 0.10), and corroborating the visual impression from Figure 4c that the schooling effect is unusually large and persistent relative to placebo units.

\bigskip

By contrast, for life expectancy (\textit{panel 5b}), the Cuban MSPE ratio sits near the middle of the donor distribution. Several placebo units exhibit post-/pre-MSPE ratios comparable to or larger than Cuba's, suggesting that the improvement in life expectancy, while positive, is less statistically exceptional than the effects on infant mortality and education. In terms of permutation-based inference, the implied p-value is not small, and we cannot reject the null of no effect at conventional significance levels. This weaker evidence is fully consistent with the empirical and theoretical considerations discussed above. In particular, life expectancy at birth is influenced by a wider set of factors than infant mortality and schooling and is more exposed to later macroeconomic and geopolitical shocks.

\bigskip

It bears emphasizing that these permutation-style p-values are not based on asymptotic normality or parametric distributional assumptions. They exploit the finite-sample variation across donor units and rely on a notion of exchangeability under the null. Specifically, in the absence of the Cuban reform, the treated unit's post-1961 behavior would not be systematically different from that of similar donor countries. This is precisely the environment for which synthetic control and related methods were designed.

\subsection{Robustness Checks}

\subsubsection{Leave-One-Out Analysis}

A natural concern in synthetic control applications is that the estimated counterfactual might be unduly driven by a single donor country with a large weight in the synthetic combination. If the treatment effect were highly sensitive to the inclusion of one particular donor, a skeptic could argue that our results reflect idiosyncratic features of that donor rather than the broader pattern in the comparison group. To address this concern, we conduct a leave-one-out (LOO) analysis in which we iteratively remove each donor country from the donor pool and re-estimate the synthetic control and the associated treatment effect.

\bigskip

This exercise serves two distinct purposes. First, it provides a direct diagnostic of the stability of our findings with respect to donor-pool composition. If the estimated gaps between Cuba and synthetic Cuba remain similar when high-weight donors are removed, the results cannot be attributed to any single comparison unit. Second, it probes whether the inclusion of donors that differ markedly from Cuba on observables---most notably the United States---creates artificial leverage in the synthetic control. If the effects persist when such donors are excluded, concerns about extrapolation from `inappropriate'' comparators are substantially mitigated.

\bigskip

\begin{figure}[H]
\centering
\begin{subfigure}[b]{0.32\textwidth}
\centering
\includegraphics[width=\textwidth]{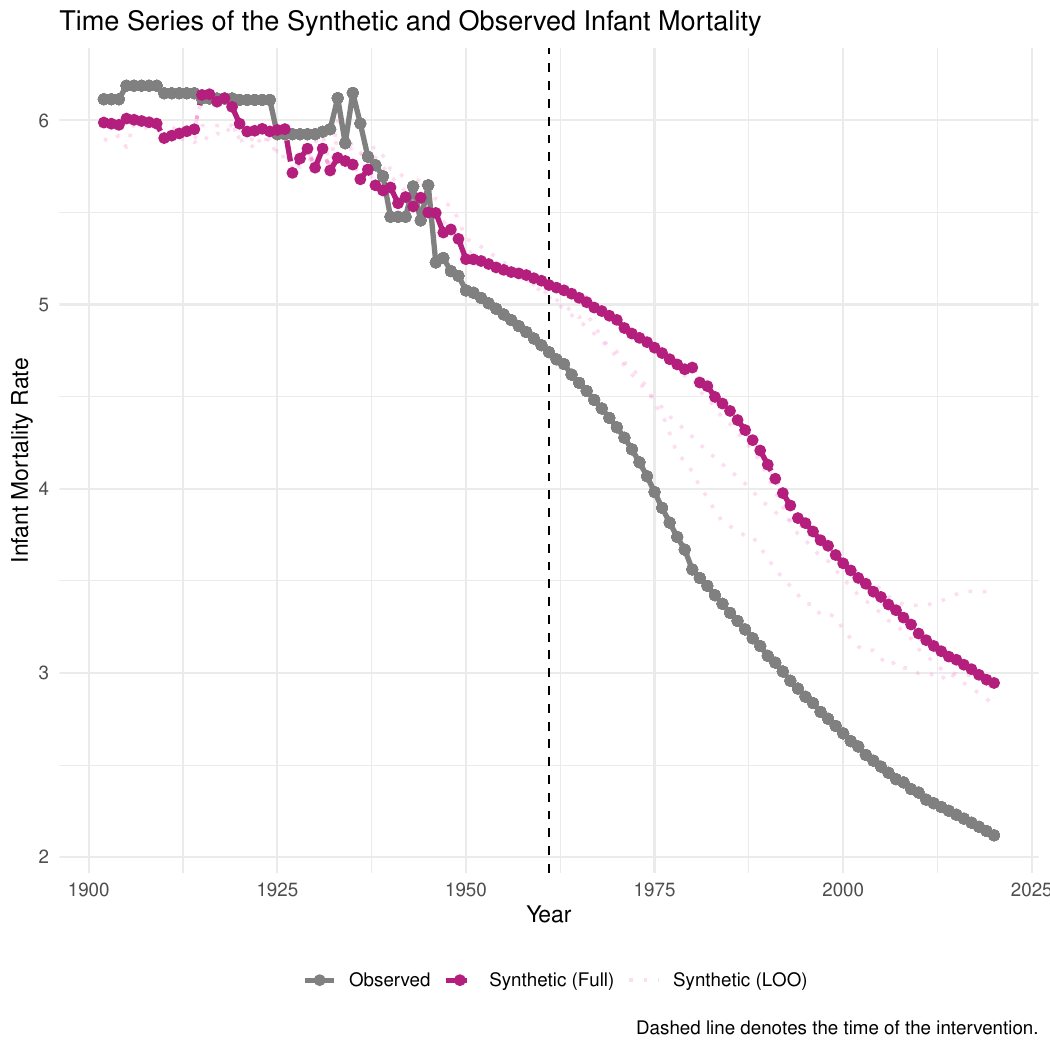}
\caption{Infant Mortality}
\label{fig:scminf_loo}
\end{subfigure}
\hfill
\begin{subfigure}[b]{0.32\textwidth}
\centering
\includegraphics[width=\textwidth]{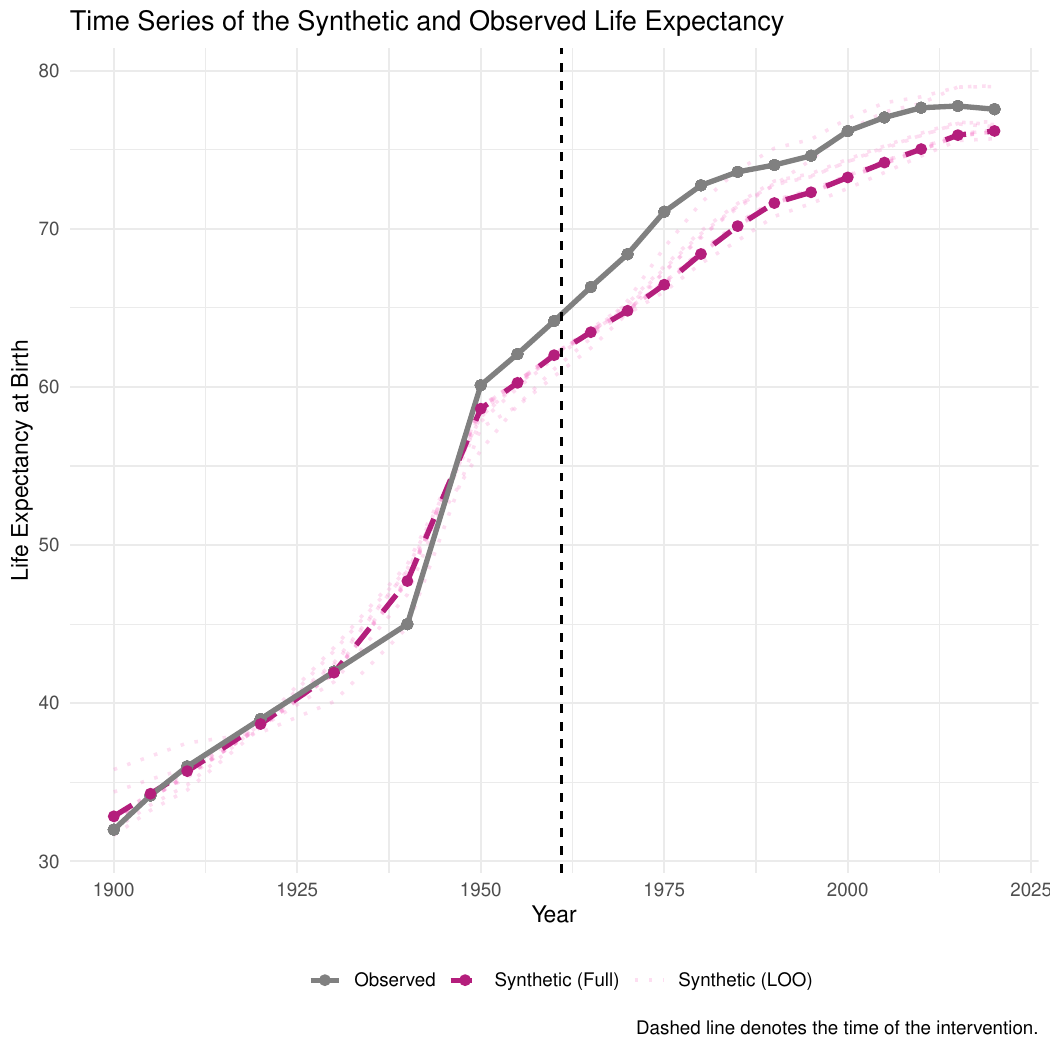}
\caption{Life Expectancy}
\label{fig:scmlife_loo}
\end{subfigure}
\hfill
\begin{subfigure}[b]{0.32\textwidth}
\centering
\includegraphics[width=\textwidth]{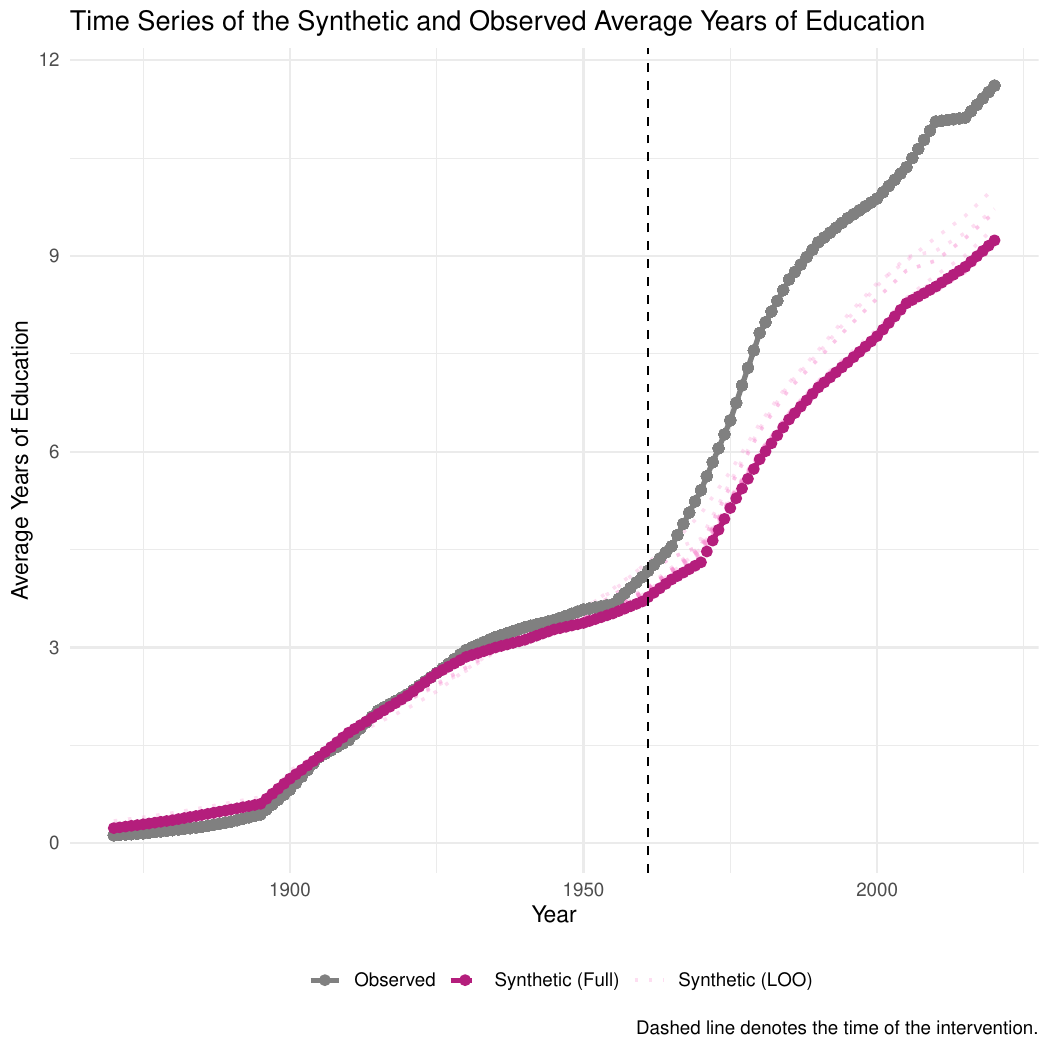}
\caption{Education}
\label{fig:scmschool_loo}
\end{subfigure}
\caption{Leave-One-Out Analysis}
\label{fig:loo_all}
\end{figure}

Figure \ref{fig:loo_all} reports the results of the leave-one-out analysis for each of our three outcomes. For each outcome, we plot three series: (i) Cuba's observed trajectory, (ii) the baseline synthetic Cuba constructed from the full donor pool, and (iii) the range of synthetic trajectories obtained when one donor country is dropped at a time. The shaded or multiple LOO lines therefore represent a collection of alternative synthetic Cubas, each based on a slightly different donor set. For infant mortality, the leave-one-out synthetic trajectories almost perfectly overlap with the baseline synthetic control over the entire pre-1961 period and remain tightly clustered in the post-reform period. The divergence between Cuba and synthetic Cuba observed in the baseline specification is preserved across essentially all LOO variants. Cuba's infant mortality falls well below all alternative synthetic trajectories after the introduction of universal health care and remains lower for the rest of the sample. This indicates that the estimated reduction in infant mortality is not an artifact of any single donor country. Rather, it reflects a robust difference between Cuba and a wide range of plausible counterfactuals drawn from the donor pool.

\bigskip

For life expectancy, the LOO synthetic trajectories again track the baseline synthetic path closely before 1961, and the post-reform pattern is very similar across specifications. Removing individual donors leads to small shifts in the level of the synthetic series but does not eliminate the post-1961 gap between Cuba and synthetic Cuba. The estimated life-expectancy effect is therefore not driven by a single high-weight donor. At the same time, the leave-one-out exercise confirms the earlier inference that the life-expectancy gap is more sensitive to later-period shocks than the infant-mortality gap, a feature that is consistent across all donor-pool variants. Lastly, the results for education exhibit a similar stability. The LOO synthetic schooling trajectories are indistinguishable from the baseline synthetic series in the pre-reform period and remain tightly bunched afterwards. In all cases, Cuba's observed schooling path diverges sharply and persistently upward from the cloud of LOO synthetics after the early 1960s. The magnitude and persistence of the schooling gap are essentially unchanged when any one donor country is removed, including the high-weight contributors Costa Rica, Panama, and the Dominican Republic. This robustness is particularly important, given that education is the outcome for which we estimate the largest proportional treatment effect.

\bigskip

A frequent concern in the literature is that including a rich, institutionally distinct country such as the United States in the donor pool might generate an implausible synthetic control that implicitly extrapolates from a combination of high-capacity and low-capacity states. The leave-one-out analysis directly addresses this issue. When the United States is excluded from the donor set and the synthetic control is re-estimated, the resulting synthetic trajectories remain extremely close to those obtained with the full donor pool, and the post-1961 gaps for all three outcomes are virtually unchanged. This finding suggests that the United States does not exert undue influence on the results and that the estimated counterfactuals are primarily driven by Latin American donors whose pre-reform trajectories closely resemble Cuba's. Overall, the leave-one-out analysis demonstrates that our conclusions are not particularly sensitive to the inclusion or exclusion of any single donor country. The pre-intervention fit remains excellent and the post-intervention divergences in infant mortality, life expectancy, and schooling remain qualitatively and quantitatively similar across all LOO specifications. Combined with the placebo and alternative-estimator evidence, this robustness to donor-pool composition reinforces the interpretation that the observed gaps reflect the causal impact of Cuba's universal health and education reforms rather than idiosyncratic features of particular comparison countries.

\subsubsection{Alternative Intervention Timing}

A key identifying assumption underlying our empirical strategy is that, absent the establishment of universal health care and the contemporaneous literacy and education reforms in the early 1960s, Cuba would have continued to follow the trajectory of its synthetic counterpart. One way to probe this assumption is to examine whether large and systematic divergences between Cuba and its synthetic peer emerge before the actual reform date when no major, comparable policy change occurred. If our results were driven by slowly evolving, unmodeled trends, we would expect to observe sizable gaps even in periods that predate the 1960s reforms.

\bigskip

To address this concern, we conduct a placebo in time exercise by backdating the intervention to 1940. We re-estimate the synthetic control for each outcome, treating 1940 as if it were the treatment year, and then recompute the post-`treatment'' gaps between Cuba and synthetic Cuba. Importantly, the donor pool and predictor set remain unchanged whilst only the assignment of the intervention date is altered. The choice of 1940 is deliberate. It is positioned well before the revolutionary period and the establishment of the National Health Service, yet sufficiently late in the sample to allow for a long pre-1940 period in which to assess fit and a substantial pseudo-post period for the placebo test.

\begin{figure}[!htbp]
\centering
\begin{subfigure}[b]{0.32\textwidth}
\centering
\includegraphics[width=\textwidth]{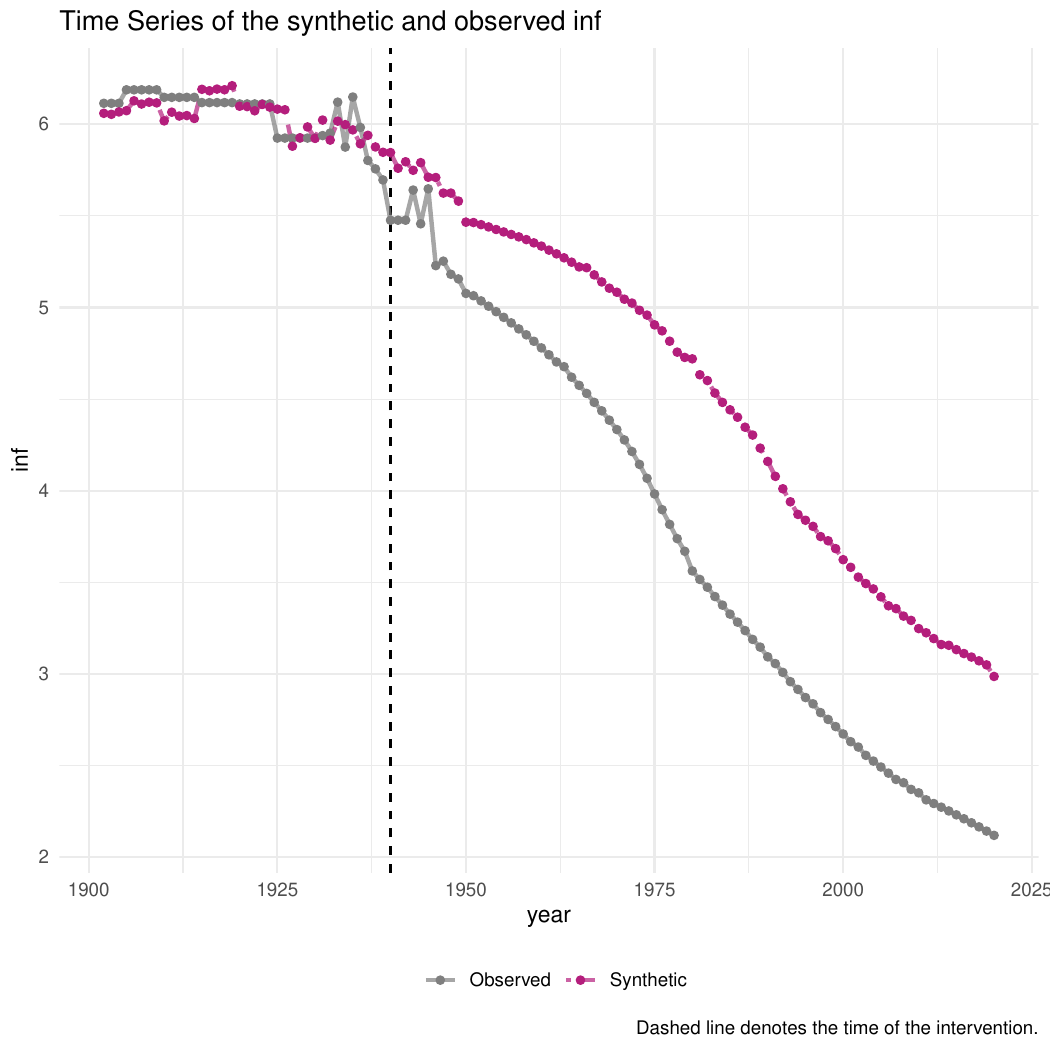}
\caption{Infant Mortality}
\label{fig:scminf_bd}
\end{subfigure}
\hfill
\begin{subfigure}[b]{0.32\textwidth}
\centering
\includegraphics[width=\textwidth]{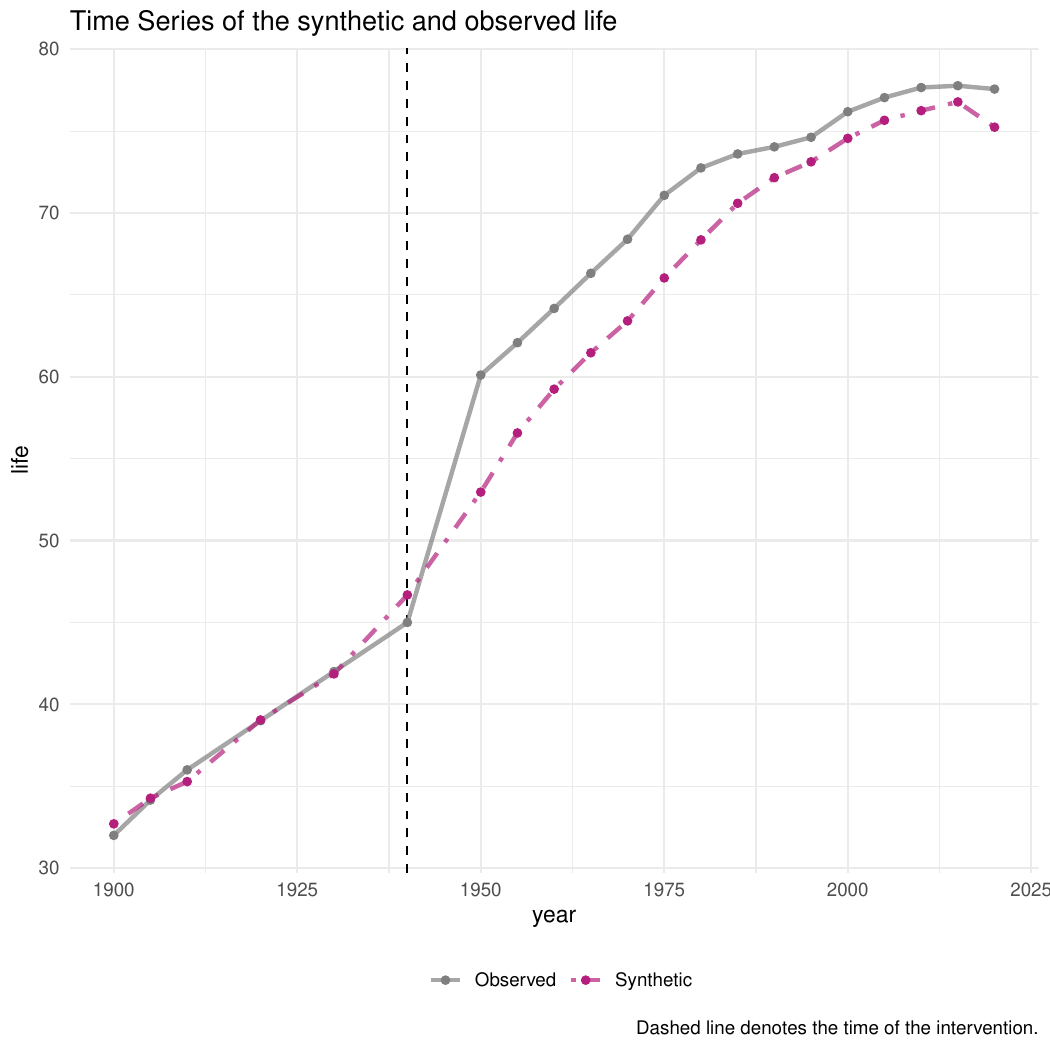}
\caption{Life Expectancy}
\label{fig:scmlife_bd}
\end{subfigure}
\hfill
\begin{subfigure}[b]{0.32\textwidth}
\centering
\includegraphics[width=\textwidth]{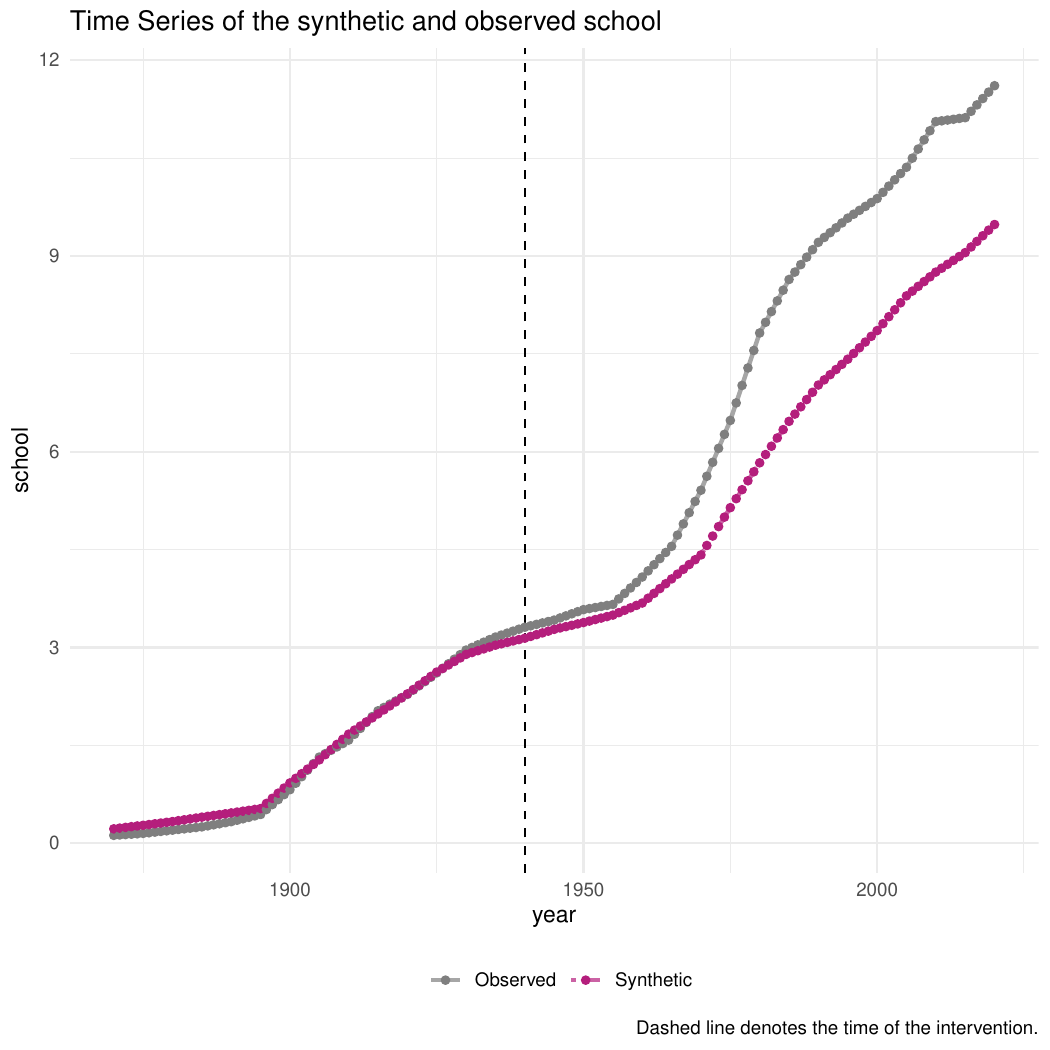}
\caption{Education}
\label{fig:scmschool_bd}
\end{subfigure}
\caption{Backdating Analysis: 1940 as Alternative Treatment Date}
\label{fig:bd_all}
\end{figure}

As a placebo test in time, we re-estimate the model using 1940 as a hypothetical intervention date. The results are reported in Figure \ref{fig:bd_all}. For infant mortality (\textit{panel 7a}), the fit between Cuba and synthetic Cuba remains tight not only up to 1940 but also well beyond it, with no persistent or monotone divergence after the placebo intervention date. Both series continue to decline gradually and in parallel, and the discrepancies that do arise are small relative to those observed after the actual reform in the early 1960s in our baseline specification. This suggests that the large post-1961 gaps documented in Figure~\ref{fig:scminf} are not simply the continuation of pre-existing differential trends that were already visible in the 1940s or 1950s.

\bigskip

For life expectancy (\textit{panel 7b}), the same conclusion holds. The synthetic control tracks the Cuban series closely before and after 1940, and the paths remain intertwined for several decades following the placebo date. There is no sign of a sharp or sustained shift in Cuba's longevity relative to synthetic Cuba around 1940 that would resemble the post-1961 divergence in our main analysis. Any deviations that emerge are modest and transient, consistent with ordinary sampling variability and small measurement differences rather than with a structural break.

\bigskip

For education (\textit{panel 7c}), the Cuban and synthetic trajectories likewise show no evidence of a discrete, post-1940 divergence. Average years of schooling in Cuba evolve very similarly to those in synthetic Cuba in the decades before the actual reforms, and there is no systematic, widening gap immediately after the placebo date. This is particularly important, because schooling is the outcome for which we find the largest proportional treatment effect in the baseline analysis. The absence of a comparable schooling divergence after 1940 reinforces the interpretation that the large, persistent gap emerging after the early 1960s is indeed associated with the joint health and literacy reforms rather than with slow-moving, pre-existing differences in education policies.

\bigskip

Taken together, the placebo-in-time exercise provides a stringent falsification test. If our baseline findings were indeed artifacts of unobserved factors that caused Cuba to diverge gradually from the donor pool irrespective of the health and education reforms, then we would expect to see sizeable and persistent gaps when we artificially declare 1940 as the treatment date. Instead, the synthetic control continues to fit Cuba well through and beyond 1940 for all three outcomes, with no pattern resembling the sharp and durable post-1961 divergences documented in our baseline results. This strengthens the causal interpretation of our results. In particular, the break in Cuba's trajectory relative to synthetic Cuba is tightly linked to the timing of the actual policy intervention and does not appear when the intervention date is shifted to a period with no comparable structural reform.

\subsection{Alternative Estimators}

The permutation-based inference described above directly addresses the question of how unusual Cuba's synthetic-control gap is relative to those of comparable countries that did not experience the reform. However, it does not provide pointwise confidence intervals in the classical sense, nor does it exploit alternative modeling strategies that can complement the baseline SCM analysis.

\bigskip

To strengthen our inferential conclusions, we therefore combine the placebo-based approach with three additional families of estimators, each with its own inferential framework. First, we implement the Augmented Synthetic Control Method of \citet{benmichael2021augmented}, which relaxes the convexity constraints of standard SCM and uses an outcome model with a ridge penalty to adjust for remaining pre-treatment discrepancies. For this estimator, we construct confidence intervals using the conformal-inference procedure developed by \citet{chernozhukov2021exact}. As reported in Table~\ref{tab:aug_synthetic_control_results} and Figure~\ref{fig:ascm_all}, the resulting intervals for infant mortality and education exclude zero, while those for life expectancy do not, mirroring the qualitative conclusions from the MSPE-ratio analysis.


\subsubsection{Augmented Synthetic Control Method}

The baseline synthetic control estimates rest on the idea that a convex combination of donor units can approximate the treated unit's pre-intervention trajectory. When pre-treatment fit is very good, as in our case, the standard SCM estimator is appealing and transparent. However, as emphasized by \citet{benmichael2021augmented}, any small residual mismatch in the pre-treatment period can translate into bias in the estimated treatment effect, especially when the post-period is long. The Augmented Synthetic Control Method (ASCM) addresses this concern by combining SCM-style weighting with an explicit outcome model that adjusts for remaining pre-treatment discrepancies, and by relaxing the pure convex-combination constraint.

\bigskip

Formally, ASCM proceeds in two steps. First, it constructs synthetic weights as in standard SCM, choosing a weighted average of donor units to minimize the pre-treatment mean squared prediction error. Second, it augments this synthetic control by fitting a regularized outcome model, typically a ridge regression of the treated unit's outcomes on donor outcomes and predictors, and then uses this model to adjust for any residual pre-treatment gaps. Conceptually, the estimator can be viewed as an SCM estimate plus a bias-correction term that exploits information about how pre-treatment discrepancies map into outcomes. This procedure retains the interpretability and design-based spirit of SCM while gaining robustness to imperfect pre-treatment balance.

\bigskip

In our context, ASCM is particularly attractive for two reasons. First, the post-treatment window is long (over six decades), so even small pre-treatment discrepancies could, in principle, cumulate into non-trivial bias if left uncorrected. Second, some donor units experience strong long-run improvements in health and education unrelated to Cuba's reform, and ASCM can partially adjust for this by leveraging the structure of pre-treatment differences. To quantify uncertainty, we construct pointwise confidence intervals using the conformal inference procedure proposed by \citet{chernozhukov2021exact}, which is specifically tailored to synthetic and related counterfactual methods and does not rely on large-sample normality.

\bigskip

\begin{figure}[!htbp]
\centering
\begin{subfigure}[b]{0.32\textwidth}
\centering
\includegraphics[width=\textwidth]{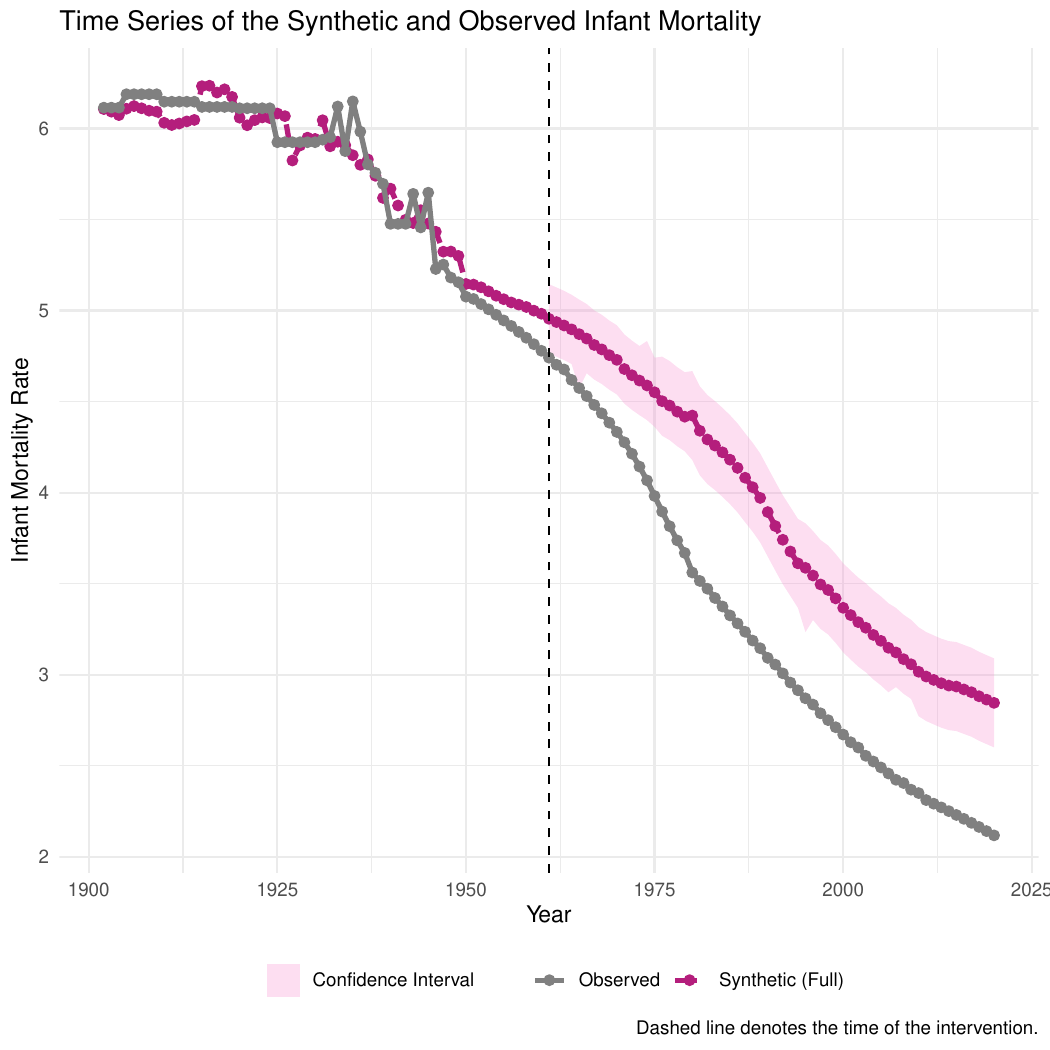}
\caption{Infant Mortality}
\label{fig:scminf_ascm}
\end{subfigure}
\hfill
\begin{subfigure}[b]{0.32\textwidth}
\centering
\includegraphics[width=\textwidth]{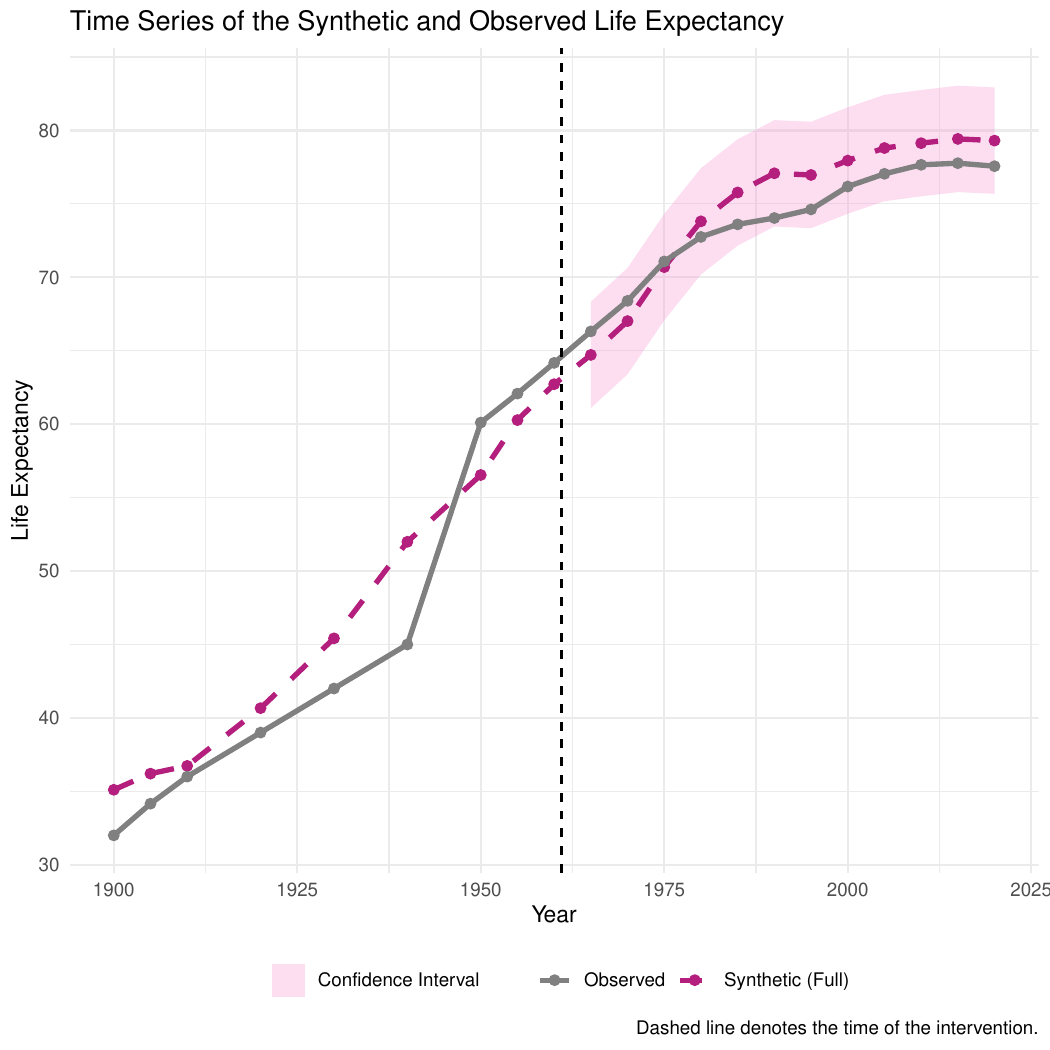}
\caption{Life Expectancy}
\label{fig:scmlife_ascm}
\end{subfigure}
\hfill
\begin{subfigure}[b]{0.32\textwidth}
\centering
\includegraphics[width=\textwidth]{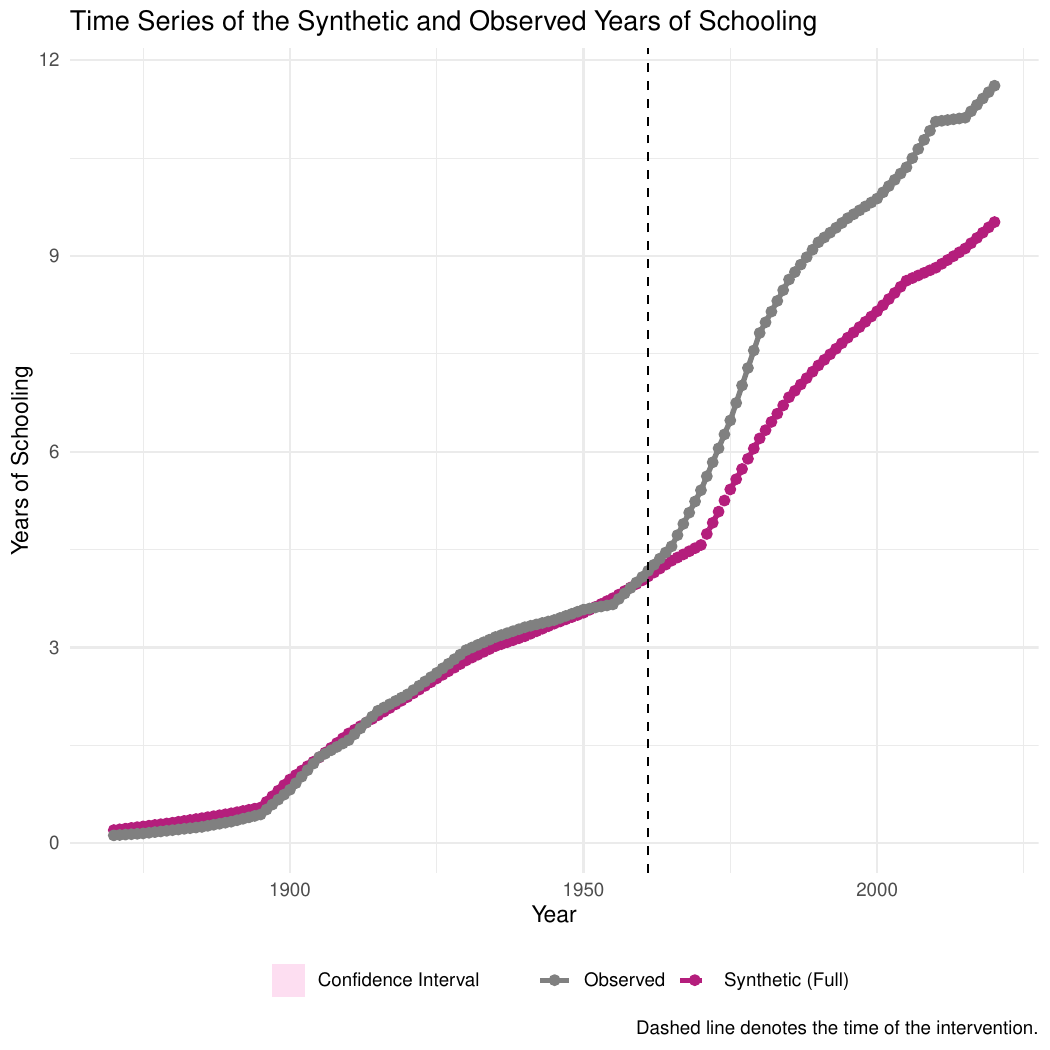}
\caption{Education}
\label{fig:scmschool_ascm}
\end{subfigure}
\caption{Augmented Synthetic Control Method Results}
\caption*{\footnotesize Note: Confidence intervals are constructed using \cite{chernozhukov2021exact}}
\label{fig:ascm_all}
\end{figure}

Table \ref{tab:aug_synthetic_control_results} reports and Figure \ref{fig:ascm_all} presents the ASCM estimates and 95 percent confidence intervals for each outcome. For infant mortality, the augmented estimate indicates an average post-treatment effect of -0.64 deaths per 1,000 live births, corresponding to an 11\% reduction relative to the synthetic counterfactual. The 95 percent confidence interval [-0.95, -0.33] excludes zero, signaling a statistically significant effect that is slightly smaller in magnitude than the baseline estimate of SCM (-0.84), but leaves the substantive conclusion unchanged. Establishment of universal health care provision is associated with a large and precisely estimated improvement in infant survival.

\bigskip

For human capital outcome, the ASCM estimate is +1.51 years of schooling, implying an 80\% increase relative to synthetic Cuba. The 95 percent confidence interval [1.12, 1.90] is entirely positive and relatively tight, confirming that the schooling effects are both large and very precisely estimated. Compared to the baseline SCM estimate of +1.84 years, the augmented estimate is somewhat more conservative but reinforces the conclusion that the reform permanently shifted Cuba onto a much higher human-capital path. The fact that the lower bound of the confidence interval still corresponds to more than a full additional year of schooling underscores the quantitative importance of this effect.

\bigskip

By contrast, the life expectancy results are more nuanced. The ASCM point estimate is -1.14 years, with a 95\% confidence interval [-3.42, 1.14] that includes zero. This stands in contrast to the baseline SCM estimate of +2.93 years. The augmented estimator, by construction, places more weight on adjusting for small pre-treatment differences and allows for extrapolation beyond the strict convex hull of donor units. In our setting, this makes the life-expectancy estimate more sensitive to donors that experienced strong post-1960 gains in longevity unrelated to Cuba's reform, as well as to modest deviations in pre-treatment trends. The resulting confidence interval, which spans both moderately negative and moderately positive values, indicates that the long-run effect on life expectancy is not statistically distinguishable from zero once this additional modeling structure is imposed.

\bigskip

We view this divergence between SCM and ASCM for life expectancy not as a contradiction but as an informative robustness check. Across the multitude of ASCM-based outcome specifications, the evidence for infant mortality and schooling is uniformly strong and statistically decisive, while the evidence for life expectancy is systematically weaker and method-dependent. The ASCM results sharpen this conclusion by showing that, under a more aggressive adjustment for pre-treatment discrepancies and donor dynamics, the life-expectancy effect becomes imprecisely estimated, whereas the infant-mortality and schooling effects remain large, robust, and precisely estimated. The pattern of the evidence is consistent with the theoretical mechanisms discussed earlier. Survival in the first year of life and accumulated years of schooling respond strongly and persistently to universal health and literacy reforms, while overall life expectancy is more exposed to subsequent macroeconomic shocks and broader demographic forces that are only partly captured by our design.

\begin{table}[ht]
\centering
\caption{Augmented Synthetic Control Results}\label{tab:aug_synthetic_control_results}
\begin{tabular}{lccc}
\toprule
 & Infant Mortality & Life Expectancy & Years of Education \\
 & (1) & (2) & (3) \\
\midrule
Augmented SCM Estimate & -0.64 & -1.14 & +1.51 \\
 & (-11\%) & (-2.4\%) & (+80\%) \\
95\% Confidence Interval & [-0.95, -0.33] & [-3.42, 1.14] & [1.12, 1.90] \\
\bottomrule
\end{tabular}
\medskip
\footnotesize \textit{Note:} Confidence intervals constructed following \citet{chernozhukov2021exact}.
\end{table}

\subsubsection{Synthetic Difference-in-Differences}

While the synthetic control method is well-suited to settings with a single treated unit and a modest donor pool, it remains fundamentally a design-based tool, with limited scope for modeling time-varying unobservables beyond what is captured through pre-treatment matching. To complement the SCM and ASCM estimates, we therefore turn to the Synthetic Difference-in-Differences (SDID) estimator proposed by \citet{arkhangelsky2021synthetic}. SDID marries the weighting intuition of synthetic control with the regression structure of difference-in-differences, jointly choosing unit weights and time weights so as to balance pre-treatment outcomes along both dimensions and then estimating a DiD-style treatment effect on the reweighted panel.

\bigskip

Conceptually, SDID relaxes some of the stringent requirements of pure synthetic control. By estimating both unit-specific and time-specific weights, it allows for richer forms of unobserved heterogeneity and time-varying confounders than a standard two-way fixed-effects model, while avoiding the strict requirement that the treated unit lie in the convex hull of the donors. In practice, this makes SDID particularly attractive in settings with long panels and potential concerns about slow-moving unobservables or mild pre-treatment misfit. To quantify uncertainty, we follow \citet{arkhangelsky2021synthetic} and compute standard errors and confidence intervals using a bootstrap that is robust to serial correlation and heteroskedasticity.

\bigskip

Table~\ref{tab:sdid_results} reports the SDID estimates for our three outcomes. For infant mortality, the SDID point estimate is -0.181 (in log points), corresponding to an 18.1\% reduction relative to the counterfactual path implied by the reweighted donors. The standard error is 0.104, yielding a 95 percent confidence interval of [-0.386, 0.022] and a p-value of 0.081. This interval narrowly straddles zero but is concentrated on negative values, indicating that the data are more consistent with a sizable improvement than with no effect. The magnitude is very much in line with the baseline SCM and ASCM results once one accounts for differences in scale and parametrization. The SDID evidence thus reinforces the conclusion that the reform led to a large reduction in infant mortality, while also recognizing that, under this more flexible specification and conservative bootstrap inference, the effect is marginally significant at the 10-percent level.

\bigskip

For education, the SDID estimate is +1.534 years of schooling, with a standard error of 0.522, a 95 percent confidence interval of [0.509, 2.558], and a p-value of 0.003. This estimate is strikingly close to the SCM and ASCM schooling effects and is statistically significant at well below the 1 percent level. The lower bound of the confidence interval still implies more than half a year of additional schooling relative to the SDID counterfactual, while the upper bound exceeds 2.5 years. These results are fully consistent with the view that the reform induced a large, structural shift in human capital accumulation that is robust to a wide range of modeling assumptions about trends, unobservables, and weighting schemes.

\bigskip

By contrast, the SDID estimate for life expectancy is modest and imprecise: +1.248 years, with a standard error of 4.277, a 95 percent confidence interval of [-9.361, 7.134], and a p-value of 0.770. Under SDID's more flexible structure, the life-expectancy effect is not statistically distinguishable from zero, and the confidence interval is wide enough to accommodate both moderate gains and moderate losses. This again echoes the pattern observed with ASCM, IFE, and matrix completion. Once we move away from the tightly constrained SCM design and allow for more general forms of unobserved heterogeneity, the data no longer support a precise estimate of the long-run effect on life expectancy.

\bigskip

\begin{figure}
    \centering
    \includegraphics[width=0.9\linewidth]{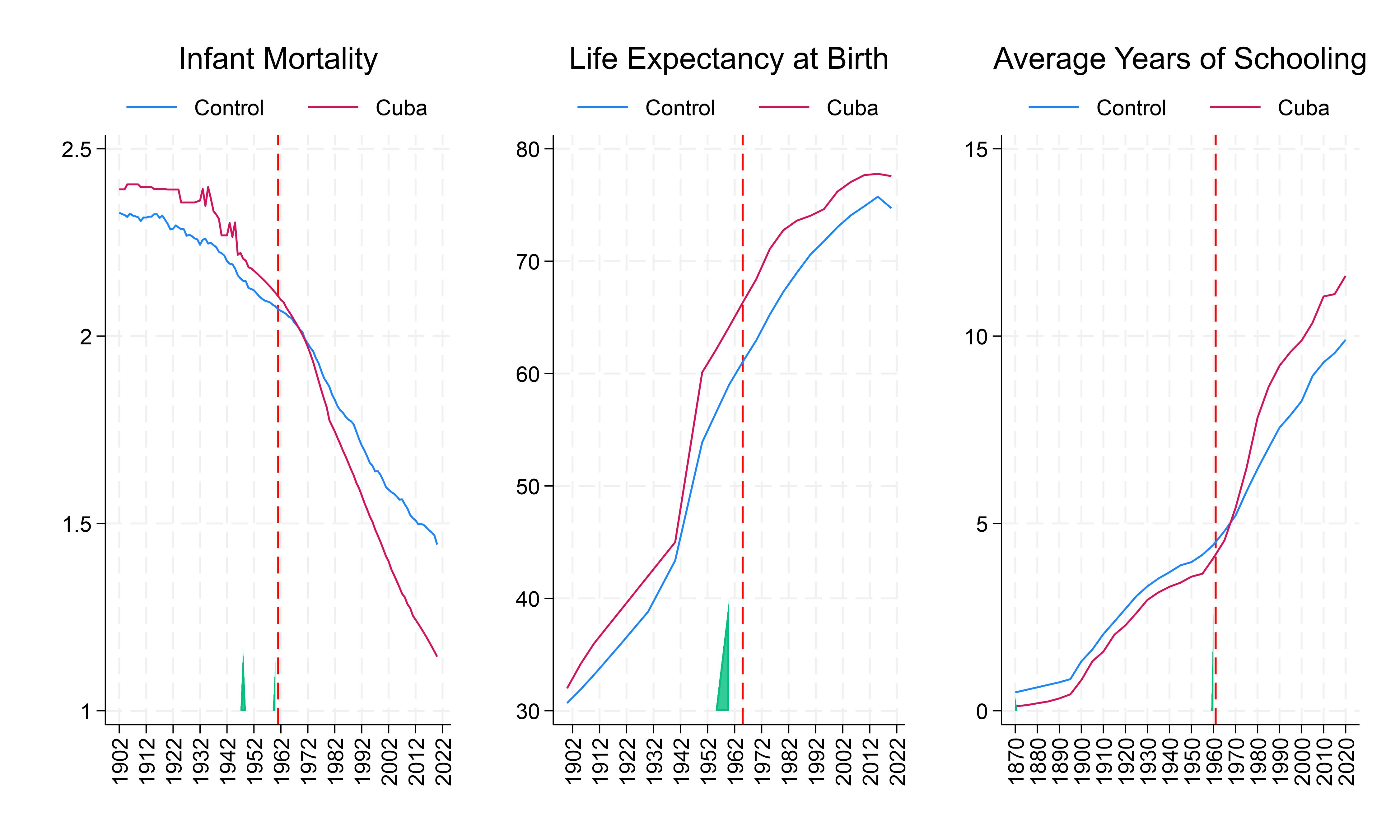}
    \caption{Synthetic Difference-in-Differences Treatment Effects Over Time}
    \label{fig:sdid_dynamic}
\end{figure}

Figure~\ref{fig:sdid_dynamic} plots the time path of the SDID treatment effects for each outcome. For infant mortality and education, the dynamic SDID estimates closely track the qualitative pattern found in the SCM results. A rapid and persistent decline in infant mortality relative to the reweighted donors, and a gradually widening and durable schooling gap. For life expectancy, the SDID path oscillates around zero with wide confidence bands, reinforcing the impression that any long-run divergence in longevity is modest relative to the sampling uncertainty under this more flexible estimator. Taken together, the SDID analysis strengthens our main conclusions in two important ways. First, it shows that the large improvements in infant survival and schooling remain when we adopt an estimator that explicitly combines the strengths of synthetic control and difference-in-differences and when we use conservative, bootstrap-based inference. Second, it confirms that the evidence for an effect on life expectancy is inherently weaker and more sensitive to modeling choices, a feature that is consistent with the theoretical mechanisms and the historical shocks, such as the `Special Period'', that affected Cuba's mortality profile in the later decades of the sample.

\begin{table}[h]
    \centering
    \caption{Synthetic Difference-in-Differences Results}\label{tab:sdid_results}
    \begin{tabular}{lccc}
        \toprule
        & \textbf{Infant Mortality} & \textbf{Life Expectancy} & \textbf{Years of Education} \\
        & (1) & (2) & (3) \\
        \midrule
        SDID Estimate & -0.181 & +1.248 & +1.534 \\
        Standard Error & (0.104) & (4.277) & (0.522) \\
        95\% Confidence Interval & [-0.386, 0.022] & [-9.361, 7.134] & [0.509, 2.558] \\
        P-value & 0.081 & 0.770 & 0.003 \\
        \bottomrule
    \end{tabular}
    \medskip
    \footnotesize \textit{Note}: Treatment effects computed for 1870-2022. Standard errors adjusted for serial correlation and heteroskedasticity using 1,000 bootstrap iterations.
\end{table}


\subsubsection{Interactive Fixed Effects and Matrix Completion}

The synthetic control and synthetic DiD estimators are designed to approximate the treated unit's counterfactual path by reweighting donor units and, in the case of SDID, time periods. A complementary approach is to model the entire outcome panel using a low-dimensional factor structure, thereby capturing common shocks and heterogeneous unit responses in a flexible way. In this subsection, we implement two such estimators, (i) the Interactive Fixed Effects (IFE) method of \citet{xu2017generalized} and (ii) the matrix completion approach of \citet{athey2021matrix}. These methods treat the untreated outcomes for the treated unit as missing entries in a partially observed matrix and impute them using a low-rank representation of the data. The associated framework generalizes standard two-way fixed effects and is well suited to settings in which unobserved confounders vary over time and have differential impacts across countries.

\bigskip

The matrix completion approach of \citet{athey2021matrix} adopts a closely related perspective. It views the outcome matrix $Y$ as approximately low-rank and estimates the missing counterfactual entries by solving a nuclear norm-regularized optimization problem, effectively performing a regularized principal components decomposition on the observed entries. Different values of the regularization parameter (denoted by lambda) trade off bias and variance: smaller values of lambda allow a more flexible fit (lower bias, potentially higher variance), while larger values of lambda induce stronger shrinkage toward a low-rank structure (higher bias, lower variance). Uncertainty is quantified via standard errors derived from the asymptotic distribution of the estimated treatment effects under the factor-model assumptions.

\bigskip

Table~\ref{tab:healthcare_estimates} reports the IFE estimates (Panel A) and the matrix completion estimates for two values of the regularization parameter (Panels B and C). For infant mortality, all estimators deliver remarkably consistent results. The IFE estimate is $-0.247$ (log points) with a standard error of $0.066$ and a $p$-value of $0.000$, implying a roughly 25 percent reduction in infant mortality relative to the counterfactual path implied by the factor model. The matrix completion estimates are $-0.294$ (s.e.\ $0.039$, $p = 0.000$) for lambda = 0.003 and $-0.292$ (s.e.\ $0.040$, $p = 0.000$) for lambda = 0.01, corresponding to reductions on the order of 29 percent. The similarity of the estimates across different regularization levels indicates that the infant mortality effect is not an artifact of overfitting or of a particular choice of the penalty parameter. These results are fully aligned with the SCM, ASCM, and SDID findings and provide strong additional evidence that the reform induced a large, statistically precise, and robust improvement in infant survival.

\bigskip

For education, the IFE and matrix completion estimates are likewise large and precisely estimated. The IFE model yields an effect of $+2.006$ years of schooling (s.e.\ $0.357$, $p = 0.000$), while the matrix completion estimates are $+1.532$ (s.e.\ $0.117$, $p = 0.000$) for lambda = 0.003 and $+1.799$ (s.e.\ $0.307$, $p = 0.000$) for lambda = 0.01. All three estimates imply gains of roughly 1.5-2 years of schooling relative to the imputed counterfactual, closely matching the magnitudes obtained under SCM, ASCM, and SDID. The very small standard errors reflect the fact that, under the low-rank factor structure, the information in the full panel is highly informative about the treated unit's counterfactual path. Taken together, these results leave little doubt that the reform produced a quantitatively large and statistically decisive upward shift in the long-run human-capital trajectory.

\bigskip

The life expectancy results once again display a different pattern. Under IFE, the estimated effect is $+1.390$ years with a standard error of $3.056$ and a $p$-value of $0.681$; under matrix completion, the estimates are $+1.447$ (s.e.\ $1.898$, $p = 0.446$) for lambda = 0.003 and $+1.145$ (s.e.\ $1.360$, $p = 0.401$) for lambda = 0.01. In all cases, the point estimates are positive but modest in size, and the confidence intervals are wide and comfortably include zero. This mirrors the ASCM and SDID findings and reinforces the conclusion that the long-run effect on life expectancy, while plausibly positive, is not estimated with precision once we allow for rich time-varying unobservables and impose a global low-rank structure on the panel.

\begin{table}[htbp]
    \centering
    \caption{Alternative Estimators: Interactive Fixed Effects and Matrix Completion}
    \label{tab:healthcare_estimates}
    \resizebox{\textwidth}{!}{
    \begin{tabular}{lccc}
        \toprule
        & \textbf{Infant Mortality} & \textbf{Life Expectancy} & \textbf{Years of Education} \\
        & (1) & (2) & (3) \\
        \midrule
        \textit{Panel A: Interactive Fixed Effects} & & & \\
        Estimate & -0.247 & +1.390 & +2.006 \\
        Standard Error & (0.066) & (3.056) & (0.357) \\
        P-value & 0.000 & 0.681 & 0.000 \\
        \midrule
        \textit{Panel B: Matrix Completion ($\lambda=0.003$)} & & & \\
        Estimate & -0.294 & +1.447 & +1.532 \\
        Standard Error & (0.039) & (1.898) & (0.117) \\
        P-value & 0.000 & 0.446 & 0.000 \\
        \midrule
        \textit{Panel C: Matrix Completion ($\lambda=0.01$)} & & & \\
        Estimate & -0.292 & +1.145 & +1.799 \\
        Standard Error & (0.040) & (1.360) & (0.307) \\
        P-value & 0.000 & 0.401 & 0.000 \\
        \bottomrule
    \end{tabular}
    }
\end{table}


We view the IFE and matrix completion results as an important complement to the design-based SCM evidence. Across these factor-model estimators, the direction and magnitude of the infant mortality and schooling effects are remarkably stable, and the associated $p$-values are effectively zero at conventional significance levels. By contrast, the life expectancy effect is consistently imprecise, with confidence intervals that straddle zero across all specifications. The joint message from these methods is clear: under a wide range of plausible assumptions about unobserved heterogeneity and the structure of the outcome matrix, the data strongly support large and persistent gains in infant survival and human capital and provide weaker, more ambiguous evidence of an effect on overall longevity. This pattern is precisely in line with the mechanisms we emphasized earlier---strong and durable impacts on early-life survival and schooling, and more shock-sensitive impacts on life expectancy over a period marked by profound macroeconomic and geopolitical disruptions.

\subsubsection{Comparison with U.S. States}

Up to this point, our donor pool has consisted of former European colonies in the Americas at the country level. A natural concern is that this set, while historically and institutionally comparable to Cuba, may still be too narrow to assess the external validity of our findings. To address this, we exploit the federal structure of the United States and extend the donor pool to include U.S. states as additional comparison units for infant mortality and life expectancy, where long-run state-level data are available. This yields a donor set of 44 U.S. states and 5 Latin American countries observed between 1902 and 2022.

\bigskip

Using U.S. states rather than the United States as a single aggregate has several advantages. First, it allows us to exploit substantial \emph{within-country heterogeneity} in income levels, racial and ethnic composition, disease environments, and health system performance. Historically, several Southern and border states exhibited income, mortality, and public-health profiles closer to those of middle-income Latin American countries than to the national U.S. average. Treating the United States as a single unit would compress this heterogeneity into a single high-income observation that is, in many dimensions, an outlier. By contrast, working at the state level lets the SCM-type procedures form counterfactuals from states whose pre-intervention trajectories more closely resemble Cuba's, thereby improving the plausibility of the comparison.

\bigskip

Second, the state-level perspective aligns more naturally with the logic of our identification strategy. Health policy in the United States is shaped by a combination of federal and state decisions, particularly before the advent of Medicare and Medicaid and even more so before the large expansions of public insurance in the late twentieth century. Yet, despite these differences in policy responsibility and health-system organization, \emph{no U.S. state} adopted a universal, free-at-point-of-use national health service comparable to Cuba's. This means that, while institutional details differ, U.S. states remain valid controls in the sense that they did not experience the kind of comprehensive, centrally planned health-care expansion that we study in Cuba. Including states therefore enlarges the donor pool without introducing treated units on the control side.

\bigskip

Third, from a methodological standpoint, using U.S. states mitigates the concern that our baseline results might be driven by the presence of a single large, developed comparator (the United States as a whole) whose pre-treatment weight in the synthetic control could be interpreted as extrapolation from an implausibly dissimilar unit. At the state level, the same population is decomposed into 44 separate observations with diverse trajectories and covariate profiles. The synthetic control can then select those states whose pre-1961 mortality and longevity dynamics most closely track Cuba's, while effectively downweighting states that are far from Cuba in terms of outcomes and predictors.

\bigskip

\begin{figure}[H]
    \centering
    \includegraphics[width=0.9\linewidth]{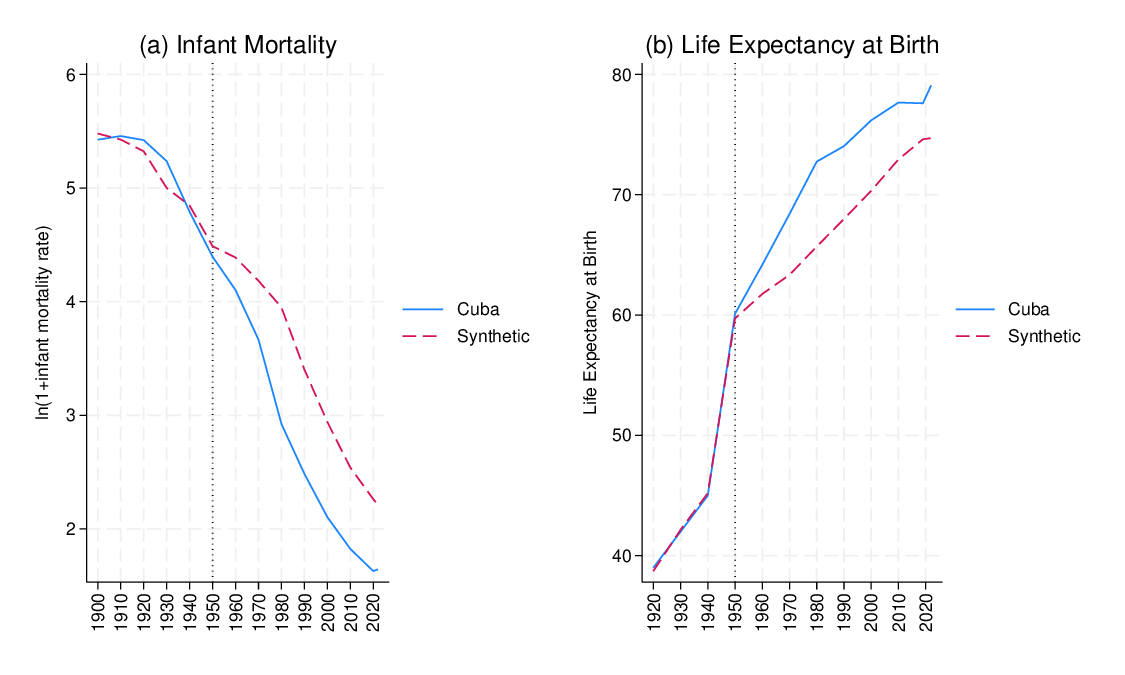}
    \caption{Long-term effect of the comprehensive national health care on life expectancy and human capital using a donor pool of US states and Latin American countries, 1902-2022}
    \label{fig:us}
\end{figure}

We re-estimate the treatment effects using this expanded donor pool and report the results in \ref{fig:us}. The findings are closely aligned with, and in some respects even stronger than, our baseline estimates. Using U.S. states and Latin American countries as donors, we find an average reduction in infant mortality of approximately 22\% (with a $p$-value of 0.021) and an improvement in life expectancy at birth of 4.8 years (with a $p$-value of 0.065).  The infant-mortality effect remains large, precisely estimated, and statistically significant at conventional levels, while the life-expectancy effect is somewhat larger in magnitude than in the baseline country-level specification and marginally significant at the 10 percent level.

\bigskip

The fact that these results emerge from a donor pool that includes a much wider range of socioeconomic and epidemiological environments, and that treats U.S. states rather than the national United States as potential controls, strengthens both the robustness and external validity of our conclusions. The Cuban reforms generate sizeable gains in infant survival and notable improvements in longevity even when the comparison set is expanded to encompass a large set of subnational units from a high-income federal country. Far from being an artifact of comparisons with a handful of Latin American countries, the estimated effects persist, and, for infant mortality, remain quantitatively similar, when we allow synthetic Cuba to be constructed from a much richer cross-section of potential counterfactual trajectories.

In the expanded donor pool that includes U.S. states, the synthetic control for infant mortality is a convex combination of El Salvador (56 percent), Connecticut (28 percent), and Mexico (15 percent). For life expectancy, the synthetic Cuba is constructed from Paraguay (77 percent), the District of Columbia (19 percent), Uruguay (4 percent), and South Dakota (less than 1 percent). These weights reflect the fact that the algorithm is matching Cuba's pre-intervention outcomes and trajectories, not institutional labels. It selects those units whose mortality and longevity paths, together with key covariates, best replicate Cuba before the introduction of universal health care.

\bigskip

At first sight, the presence of Connecticut and the District of Columbia in these combinations may appear surprising, but both play a natural role in matching Cuba's position prior to the revolution. In the infant-mortality specification, El Salvador and Mexico provide the higher pre-1960 mortality levels and tropical disease environment characteristic of middle-income Latin America, while Connecticut contributes a low-mortality but steadily improving public-health system whose downward trend in infant mortality closely mirrors Cuba's pre-reform decline. Like pre-revolutionary Cuba, Connecticut combined relatively advanced medical infrastructure in urban centers with pockets of deprivation among poorer and immigrant communities, generating an intermediate profile that lies between the very low mortality of the healthiest U.S. states and the much higher rates observed in the poorest Latin American countries. The resulting synthetic unit thus reproduces both Cuba's level of infant mortality and its slope of decline over 1900-1960.

\bigskip

A similar logic applies to life expectancy and the role of the District of Columbia. Paraguay and Uruguay anchor the synthetic control in the Latin-American income and demographic range, while the District of Columbia and, to a much smaller extent, South Dakota capture features of a small, highly urban jurisdiction with marked inequalities and large mid-century gains in longevity driven by improvements in basic public-health infrastructure. This yields a synthetic path that matches Cuba's life-expectancy level and its `late but rapid'' rise relative to richer countries. Moreover, our leave-one-out analysis shows that dropping any single donor---including Connecticut or the District of Columbia---leaves the estimated treatment effects virtually unchanged, indicating that no individual U.S. state drives the results. As an additional exercise, Appendix~\ref{app:external} compares the observed life-expectancy trajectory of Cuban-Americans in Florida with a counterfactual constructed for the Havana pre-revolutionary elite, illustrating that the bundle of universal care plus a richer socioeconomic environment generates a more rapid longevity ascent than migration to the United States alone.

\subsection{Summary}

Taken together, the empirical results paint a coherent and robust picture of the long-term consequences of Cuba's universal health-care and education reforms. The baseline synthetic control analysis shows that, relative to a carefully constructed counterfactual, the introduction of the National Health Service and the contemporaneous literacy and education policies is associated with a large and persistent reduction in infant mortality, a substantial though more modest improvement in life expectancy, and a dramatic and enduring rise in average years of schooling. The timing of these divergences closely tracks the policy intervention. Pre-treatment fit is extremely tight for all three outcomes, while sizeable gaps emerge only after the early 1960s and widen or stabilize in ways that are consistent with the mechanisms laid out in Section~\ref{sec:theory}.

\bigskip

Our inference procedures support a causal interpretation of these patterns. Placebo tests in space and post-/pre-intervention MSPE ratios indicate that the gaps for Cuba in infant mortality and education are unusually large compared with those of donor countries reweighted as if they had been treated, yielding permutation-style $p$-values below conventional thresholds, whereas the life-expectancy gap is less exceptional. Augmented synthetic control estimates, combined with conformal confidence intervals, confirm statistically and economically significant effects on infant mortality and schooling but deliver a smaller and statistically inconclusive effect on life expectancy. Synthetic difference-in-differences, interactive fixed effects, and matrix completion---each relying on distinct modeling assumptions about unobserved heterogeneity and factor structures---converge on the same conclusion: the data provide strong and precise evidence of large improvements in infant survival and human capital, and weaker, method-dependent evidence of gains in overall longevity.

\bigskip

A broad set of robustness checks further strengthens these findings. When we backdate the intervention to 1940, no comparable divergence appears, indicating that the post-1961 gaps are not the continuation of pre-existing differential trends. Leave-one-out exercises show that the estimated effects are insensitive to the removal of any individual donor, including high-weight units such as the United States or particular Latin American countries. Expanding the donor pool to include U.S. states yields synthetic controls that match Cuba's pre-reform trajectories closely and produce treatment effects on infant mortality and life expectancy that are of similar or even larger magnitude than in the baseline specification, suggesting that our results are not an artifact of a narrow or idiosyncratic comparison group.

\bigskip

In sum, across a wide range of estimators, inferential strategies, donor pools, and specification choices, the evidence consistently points to a major and enduring impact of Cuba's post-revolution health and education reforms on early-life survival and human capital accumulation. The effect on life expectancy at birth is positive but less precisely estimated, reflecting both the broader set of determinants of adult mortality and the influence of subsequent macroeconomic shocks. The combination of design-based and model-based approaches thus delivers a clear message. Universal health care, when embedded in a broader state-building project that expands literacy and schooling, can permanently shift a country onto a higher path of infant survival and educational attainment.

\section{Conclusion}\label{sec:conclusion}

In this paper, we have examined the long-run consequences of a large-scale expansion of universal health care for public health and human capital. We exploit the creation of Cuba's National Health Service in 1961, implemented alongside an ambitious literacy and education drive, as a quasi-experimental source of variation in health and schooling outcomes. Using newly assembled historical series on infant mortality, life expectancy at birth, and average years of education for a panel of former European colonies in the Americas, we construct synthetic control counterfactuals and apply a range of modern panel estimators to quantify the long-term effects of these reforms.

\bigskip

Our main finding is that the introduction of universal health care, embedded in a broader state-building project, is associated with a large and persistent improvement in early-life survival and a dramatic upward shift in the trajectory of human capital accumulation. Relative to synthetic Cuba, infant mortality falls substantially and the gap remains open for more than half a century, while average years of schooling rise by roughly between 1.5 and 2 years and continue to diverge over time. These effects are robust across a wide array of specifications: baseline synthetic control, augmented synthetic control, synthetic difference-in-differences, interactive fixed effects, matrix completion, and donor pools that include both Latin American countries and U.S. states. In every case, the estimated impacts on infant mortality and schooling are large in magnitude and precisely estimated.

\bigskip

By contrast, the evidence for a long-run effect on life expectancy is more nuanced. Baseline synthetic control estimates point to a positive and economically meaningful increase in longevity, but once we move to estimators that more flexibly model unobserved heterogeneity and long-run trends, the life-expectancy effect becomes smaller and imprecisely estimated. This pattern is consistent with the idea that life expectancy at birth is influenced by a broader set of forces---including subsequent macroeconomic and geopolitical shocks---than infant mortality and schooling, and that health-system reforms, while important, are only one determinant of long-run longevity.

\bigskip

Taken together, our results suggest that universal health care, when implemented as part of a comprehensive project that expands literacy, schooling, and state capacity, can permanently shift a country onto a higher path of infant survival and educational attainment. The Cuban reforms did not simply generate a one-off improvement in health indicators. Instead, they altered the environment in which families make human capital investments and in which the state delivers core public goods. The strong and persistent gains in infant mortality and schooling that we document provide empirical support for theories that emphasize the complementarity between early-life health, education, and state capacity, and they underscore the potential for large public-health investments to have long-lived effects on human capital and development.

\bigskip

At the same time, our analysis highlights important limitations and avenues for future work. The Cuban experience combines health, literacy, and education reforms in a bundled institutional shock, making it difficult to fully disentangle the contribution of each component. Moreover, our focus on aggregate outcomes leaves open questions about distributional effects, the fiscal and political sustainability of universal health systems, and the external validity of our findings in settings with different institutional and macroeconomic conditions. Addressing these issues will require new data and complementary research designs. Nonetheless, the Cuban case provides a rare, historically grounded example in which a low- to middle-income country undertook a radical expansion of health and education services, and the evidence presented here indicates that such reforms can generate profound and enduring improvements in both public health and human capital.

\appendix

\section{Extended Historical and Policy Background}\label{app:history}

This appendix provides a more detailed account of the historical and institutional setting summarized in Section~\ref{history}.

\subsection{The Cuban health paradox}
Cuba's health-care system has attracted scholarly and policy attention since the 1959 revolution. The Cuban health paradox refers to the country's attainment of public-health outcomes comparable to advanced economies despite limited income and an external embargo \citep{SpiegelYassi2004Cuba,Evans2008Cuba}. Detailed historical treatments are available in \citet{cooper2006health} and \citet{lamrani2013cuba}, among others.

\subsection{Pre-revolutionary health conditions}
Before 1959, the Cuban health-care system was characterized by deep inequalities between affluent urban areas and deprived rural areas. Despite scientific advances in epidemiology by Cuban scientists, the provision of health care was concentrated in Havana and the principal cities and provided almost exclusively through private clinics \citep{danielson1979cuban,keck2012curious}.\footnote{One of the best-known pre-revolutionary scientific advances is Carlos J. Finlay's discovery of the mosquito-borne transmission of yellow fever (1881), confirmed by the U.S.\ Yellow Fever Commission in 1900. Finlay's work led to the eradication of yellow fever in 1901, facilitated the completion of the Panama Canal between 1904 and 1914, and informed modern mosquito-borne disease control \citep{barnet1915finlay,chavescarballo2005finlay}.} The rural areas had a single hospital and bore the burden of widespread tuberculosis, malaria, parasitic infections, and child malnutrition.

A pervasive urban--rural gap was reinforced by the shortage of well-trained doctors, widespread illiteracy, and high rates of infant mortality. A report by \citet{truslow1951report} shows that only eleven percent of farm workers consumed milk while the rural infant-mortality rate stood at 100 per 1{,}000 live births. Rural oral health was also poor \citep{watson2022oral}, exacerbated by excessive sugar-cane consumption. Coupled with child malnutrition and widespread prostitution \citep{miller2003trading,sippial2013prostitution,kurlansky2017havana}, the public-health disparities between affluent and poor strata were severe. Pre-revolutionary life expectancy at birth has been estimated at 60--62 years in major cities and 45--50 years in rural areas, an urban--rural gap of 10--15 years. Around 80 percent of children in the countryside had intestinal parasites and 60 percent of the population was chronically malnourished \citep{mcguire2005cuba,gorry2012cuba,delgadolegon2018cuba}. These inequalities were further reinforced by the widespread corruption and authoritarian repression of the Batista regime (1952--1959).

\subsection{The post-1959 transformation}
After 1959, the Cuban health system underwent a complete transformation into a universal, state-funded, community-based health care.\footnote{The transformation, built on the principles of equity, solidarity and prevention, was the subject of Che Guevara's 1960 speech at the Pan-American Health Organization conference \citep{guevara1969speech}.} The establishment of the \textit{Sistema Nacional de Salud} in 1961 marked a formal transition to a comprehensive, universal system. The right to health care was enshrined in the constitution; hospitals and clinics were nationalized; preventive care was massively expanded into rural areas;\footnote{One of the most widely recognized expansions of preventive care is the Family Doctor Program, inaugurated in 1984, which embedded primary-care providers within the family and community environment.} and medical-school enrollment rose dramatically, eventually generating more doctors per capita than most advanced economies. \citet{DeVos2005CubaHealth} characterizes the system's development in four stages: foundation (1959--70), consolidation (1970--79), expansion of family medicine (1980--90), and reform (1990 onward).

\subsection{Scale of health investment and outcomes}
Cuba has consistently devoted a large share of GDP to health care, ranging from 6 percent of GDP (roughly one quarter of the national budget) in the early years to 9, 11, and 14 percent in 2010, 2015, and 2021 respectively \citep{sixto2002cuba,MesaLago2002CubaEconomy,anuario2021salud}. Infant mortality fell from roughly 60 deaths per 1{,}000 live births in the 1950s to around 4 today---comparable to high-income countries. Life expectancy at birth rose from 59 years in 1959 to roughly 80 years in 2023. Cuba achieved the elimination or containment of malaria (1967), dengue fever (1981), polio (1982), measles (1993), and mother-to-child transmission of HIV and syphilis (2015).

\subsection{Medical diplomacy and biotechnology}
Cuba's international medical diplomacy includes its first humanitarian mission in Algeria (1963), Operation Miracle (2004), West African Ebola missions (2014), and the dispatch of 38 brigades and 2{,}544 health-care professionals to 39 countries during the COVID-19 pandemic. Cuba treated over 25{,}000 children from Ukraine, Belarus, and Russia after the Chernobyl disaster, and established the \textit{Escuela Latinoamericana de Medicina}, which has trained more than 30{,}000 doctors from under-served communities.\footnote{Students primarily come from Africa, Latin America, the Caribbean, Asia, and the United States; graduates commit to public service in low-income regions.} Despite the U.S.\ embargo and the resulting health and economic costs \citep{garfield1997impact,barry2000effect,gordon2016embargo}, Cuba developed the Meningitis B vaccine \citep{randall2017revolution,ochoa2018meningitis,sierragonzalez2019meningitis}, a recombinant Hepatitis B vaccine that drove incidence from 10--15 to under 1 per 100{,}000 \citep{ivon2017hepatitis},\footnote{The Hepatitis B vaccination program proceeded in two stages: targeted (infants born to infected mothers, healthcare workers) and then universal.} the therapeutic lung-cancer vaccine CIMAvax-EGF \citep{crombet2015cimavax,tagliamento2018cimavax}, and the Abdala and Soberana COVID-19 vaccines \citep{masbermejo2022covid}.

\subsection{Legal and demographic context}
In 1965 Cuba became the first Latin American country to legalize abortion.\footnote{Other Latin American countries legalized abortion in 2012 (Uruguay) and 2020 (Argentina). See \citet{kulczycki2011abortion}.} In 2022, Cuba adopted a family code widely regarded as among the most progressive in Latin America \citep{adler2024,adler2023cuba}.\footnote{The 2022 family code legalized same-sex marriage and child adoption for same-sex couples, permitted altruistic surrogacy, recognized a wider range of family structures, prioritized children's rights, and prohibited discrimination on the basis of sexual orientation and gender identity.} Earlier, Cuba had been regarded as one of the best developing countries for prospective motherhood based on the Save the Children index.\footnote{\textit{Informe Sobre El Estado Mundial de las Madres 2011.} Madrid: Save the Children.}

\subsection{The post-1959 reforms in detail}
The post-1959 health-care transformation rested on several large-scale investments \citep{conover1980cuba,boffey1978cuba}. Health care was declared a human right, hospitals were brought under state control, and the Cuban government built hundreds of new hospitals and clinics across the island. In the early 1960s, the government introduced a decentralized polyclinic model providing all-in-one primary care, prevention, and specialized services across neighborhoods \citep{hirschfeld2017health}. From 1961 onward, medical education was free, and mass immunization, child-nutrition, sanitation, and prenatal-care programs were rolled out \citep{bodenstein2010cuba}. From the 1970s onward, Cuba sent doctors abroad as part of its medical diplomacy \citep{feinsilver1993healing,huish2013cuba}, and from the 1980s onward, the country built a substantial biotechnology sector \citep{villanueva2023biotech}.\footnote{The Finlay Institute (1991) and the Center for Genetic Engineering and Biotechnology (CIGB, 1985) led to several vaccine innovations.}

\subsection{The 1961 National Literacy Campaign}
Existing estimates indicate that the literacy rate before the revolution stood at 60--76 percent, with sharp urban--rural disparities \citep{leiner1987literacy}.\footnote{Urban literacy in 1959 was approximately 89 percent versus 59 percent in rural areas \citep{rey2021literacy}.} Over eight months in 1961, more than 250{,}000 volunteer \textit{brigadistas} were mobilized to teach reading and writing in disadvantaged rural and urban communities without electricity, running water, or paved roads. The campaign centered on the construction of new schools, training of educators, and instruction of peasants (\textit{guajiros}).\footnote{Higher-education access was simultaneously expanded through gender quotas and other affirmative-action policies. The policies were echoed by Che Guevara's speech at the University of Las Villas, in which he declared that ``days when education was a privilege of the white middle class'' had ended \citep{anderson1997che}.} By the end of the campaign, roughly 707{,}000 Cubans---about 10 percent of the population---had become literate, reducing the national illiteracy rate to 3.9 percent. By 1962, the literacy rate stood at 96 percent, and by 1981 had reached 98 percent.

\subsection{The Special Period and beyond}
During the 1990s ``Special Period'' that followed the collapse of the Soviet Union, Cuba maintained its free health-care system despite acute resource constraints, funding it through state revenues and international medical services. By maintaining steady investments in medicine, research, and public-health infrastructure, the government preserved its long-term commitment to health as a national policy.

\subsection{Bundled reforms and identification}
Our empirical strategy attributes the post-1961 divergence of Cuban outcomes from their synthetic counterparts to the joint introduction of universal health care and the simultaneous mass-literacy reform. The bundled-reform interpretation reflects two features of the Cuban experience. First, reductions in infant mortality and infectious-disease burden raise the expected returns to investing in schooling \citep{bleakley2010health}, while increased literacy facilitates better health knowledge and intergenerational transmission of human capital. Second, the post-revolutionary period also encompassed land reform, nationalization of industries, large-scale elite emigration, the U.S.\ embargo, and substantial Soviet support that ended abruptly in 1990. The schooling effects should therefore be interpreted as reflecting a joint institutional transformation in which health-care access and the eradication of illiteracy reinforced each other, while the life-expectancy results plausibly reflect the additional sensitivity of adult mortality to subsequent macroeconomic and geopolitical shocks.

\section{External Counterfactual: Cuban Elites vs.\ Cuban-Americans}\label{app:external}

The question that arises immediately the comparison of Cuba to the US states is simple and straightforward. How would a group most ethnically and demographically similar to Cuba as a whole perform in the hypothetical presence of large-scale health reforms in the United States? To answer this, we compare the observed life expectancy of Cuban Americans in Florida from 1920 to 2020 with the observed life expectancy of Havana's elite prior to the Cuban Revolution (1920-1950). In addition, we extend this comparison with a hypothetical counterfactual trajectory for the Havana elite assuming they had emigrated to the U.S. after 1950 and benefitted from a massive state-led healthcare investment similar to the one implemented in Cuba under Castro after the revolution. This scenario is contrasted with the actual health trajectory experienced by Cuban emigres in the U.S., where they did not receive such reforms.

\bigskip

The Cuban American life expectancy in Florida is constructed form a series of reports on Florida's Vital Statistics. It reflects the life expectancy at birth trajectory of Cuban emigrants, which began with a select group of Cuban exiles, largely from Havana's elite, followed by subsequent waves of migration. The Cuban American population in Florida grew steadily after 1959, and life expectancy rose accordingly due to factors such as migration selectivity, access to modern healthcare, and public health improvements in the U.S. The early Cuban exiles, primarily composed of the upper class from Havana, initially enjoyed health advantages, but their life expectancy gradually increased as they benefitted from U.S. healthcare reforms and greater access to medical care. Over the span of 100 years, from 1920 to 2020, Cuban-American life expectancy in Florida steadily increased, mirroring trends in U.S. public health improvements, particularly after the 1960s.

\bigskip

\begin{figure}
    \centering
    \includegraphics[width=0.9\linewidth]{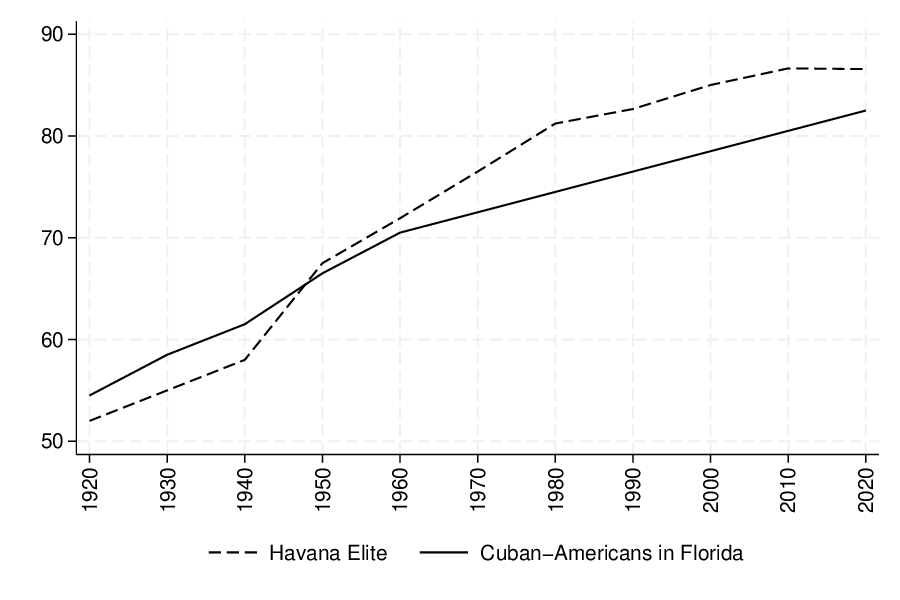}
    \caption{Observed and synthetic life expectancy at birth of Cuban-Americans in Florida}
    \label{fig:hav_elite}
\end{figure}

Figure~\ref{fig:hav_elite} depicts the observed Cuban American life expectancy at birth trajectory against the Havana elite from 1920 until 1950, followed by the hypothetical trajectory in the quasi-presence of similar life expectancy increments as we observe in Cuba under the health reform regime starting in early 1960. In contrast, the Havana elite's life expectancy from 1920 to 1950 reflects the health trajectory of Cuba's highly selected professional, business, and political class prior to the revolution. With access to private medical care, better nutrition, and superior living conditions in Havana, the elite experienced substantially higher life expectancy than the broader Cuban population. In 1950, life expectancy for the Havana elite is estimated at 65 years, reflecting the privileged position they held within Cuba's pre-revolutionary social structure. From 1920 to 1950, Havana elites enjoyed a steady increase in life expectancy, largely driven by their ability to access high-quality healthcare and other socio-economic benefits. However, this trajectory was confined to a narrow, privileged group, and did not represent the broader population, which faced challenges such as limited healthcare access, lower living standards, and higher mortality rates.

\bigskip

The counterfactual trajectory assumes that the Havana elite emigrated to the United States after 1950 and benefited from a massive state-led healthcare investment, similar to the healthcare reforms implemented in Cuba after the revolution under Fidel Castro. These reforms included universal healthcare coverage, public health programs, and massive improvements in sanitation, which contributed significantly to Cuba's health outcomes post-1959. By applying the observed life expectancy increments from Cuban-Americans in Florida between 1950 and 2020, we estimate that the Havana elite could have experienced a steady increase in life expectancy similar to the health improvements observed in the Cuban-American population in Florida, assuming they had benefited from similar health reforms.

\bigskip

From 1950 onward, this counterfactual trajectory suggests that the Havana elite would have reached a life expectancy of approximately 85.5 years by 2020, an increase of more than 20 years from the baseline in 1950. This hypothetical trajectory assumes that, had they emigrated and benefited from the state-led healthcare infrastructure in the U.S., the Havana elite would have experienced health improvements similar to those seen in Cuban-Americans in Florida, who benefitted from improvements in U.S. public health systems and access to state-funded healthcare services.

\bigskip

The gap between the observed Cuban-American life expectancy (orange line) and the hypothetical life expectancy of Havana elites (blue line) reveals significant insights into the effects of migration, healthcare access, and political disruption. While Cuban-American life expectancy in Florida increased steadily, particularly from 1960 to 1990, it remained below the hypothetical trajectory of the Havana elite. This divergence can be explained by the socioeconomic and demographic composition of the two groups.

\bigskip

First, early Cuban emigrants from Havana's elite had a higher health status due to their privileged position in Cuba, but the later waves of Cuban migrants faced challenges related to lower socioeconomic status, limited access to healthcare, and the stresses of migration. These factors resulted in slower health improvements among later Cuban-American cohorts compared to the potential trajectory of the Havana elite. And second, the observed Cuban-American trajectory shows health improvements consistent with those of other U.S. immigrant groups, but the counterfactual trajectory illustrates the maximum potential health benefit that the Havana elite could have experienced had they emigrated and benefited from state-led health reforms similar to those implemented in Cuba post-1959. This trajectory suggests that the Havana elite could have reached higher life expectancy levels than even the Cuban-American population, had they been able to benefit from healthcare reforms similar to those in Cuba under Castro.

\bigskip

This external counterfactual comparison illustrates the impact of migration and healthcare systems on long-term health outcomes. The Havana elite's hypothetical trajectory demonstrates that, had they experienced similar health care investment shock in the U.S., they would have experienced significant health benefits in the  post-1950 period. The gap between the observed and counterfactual trajectories highlights the profound and lasting impact of the Cuban revolution on health outcomes. While the Cuban-American population in Florida benefited from access to modern healthcare, the broader Cuban population in Cuba did enjoy private benefits due to the political and economic disruptions caused by the revolution but instead benefitted from large-scale public health intervention which had been largely absent in the US. The comparison highlights the complex interaction between migration, healthcare reforms, and socioeconomic conditions, and provides valuable insights into the long-term effects of the revolution on health outcomes.

\section{Extended Theoretical Mechanisms}\label{app:theory}

This appendix elaborates on the four mechanisms summarized in Section~\ref{sec:theory}.

\subsection{Early-life health, survival, and the return to human capital}
In the canonical health-capital framework, individuals derive utility both directly from health and indirectly through its effect on productivity and lifetime earnings. Health is a durable stock that depreciates over time but can be augmented by investments in medical care, nutrition, and other inputs \citep{grossman1972concept}. Universal access to preventive and curative services---prenatal care, vaccination, basic infectious-disease treatment, and community-level primary care---raises the marginal product of health investments and shifts the health production frontier outward. When such investments are rolled out in an environment characterized by high infant and child mortality, limited access to medical facilities, and a high burden of communicable disease, the immediate effect is improved early-life survival and reduced childhood morbidity. Decline in infant mortality raises the expected length of life and increases the horizon over which investments in human capital can be amortized; better childhood health improves cognitive and physical development, increasing the effectiveness of schooling and other human-capital investments. Studies of historical eradication or attenuation of parasitic diseases and major health shocks, including hookworm, malaria, and waterborne diseases, show that cohorts exposed to improved health environments in early life subsequently attain more education and enjoy higher incomes \citep{bleakley2010health,cutler2005role}. \citet{AlmondCurrie2011HumanCapital} survey similar evidence across a range of interventions.

\subsection{Health-schooling complementarities and the role of mass literacy}
A second mechanism operates through complementarities between health improvements, basic literacy, and schooling decisions. In models of human-capital formation, health, parental education, and schooling are jointly determined and mutually reinforcing \citep{BeckerTomes1979,BarroBecker1989,Heckman2007HumanCapital}. Better health raises both the capacity to learn (through cognition, attendance, and energy) and the incentive to invest in education (through longer expected horizons). When a health reform is implemented alongside a mass literacy campaign, these complementarities are amplified: literate parents are more likely to understand the benefits of education, to process written information about health and nutrition, and to support children's learning at home, while literacy among older cohorts facilitates the diffusion of health-related knowledge and improves compliance with preventive measures. Health interventions that improve childhood nutrition and reduce disease burdens often yield large gains in schooling and cognitive outcomes \citep{miguel2004worms,bhalotra2022health}. A natural implication is that the divergence in educational attainment between a treated country and its synthetic counterpart should widen gradually and persistently.

\subsection{State capacity, service delivery, and violence reduction}
A third mechanism links large-scale health and education investments to state capacity \citep{BesleyPersson2009StateCapacity,DinceccoKatz2016Capacity}. The creation of a national health service that reaches beyond major urban centers entails a significant expansion of the state's infrastructural capacity. Establishing clinics and hospitals in rural and marginalized areas, training and deploying medical professionals, and instituting surveillance and reporting systems for morbidity and mortality all deepen the state's footprint. Once such infrastructure is in place, the marginal cost of providing additional public goods, such as vaccination campaigns, literacy programs, or sanitation improvements, declines. State capacity is also closely linked to the consolidation of internal security: a state that suppresses organized crime, rural banditry, and private violence reduces violence-related mortality and creates a more predictable environment for service delivery \citep{AcemogluRobinson2006Institutions,AcemogluRobinsonSantos2013Violence}. Reduced violence and greater territorial control ease the deployment of health personnel, improve access to facilities, and lower the risks associated with travel for patients and providers. By embedding clinics, schools, and literacy brigades into local communities, the state acquires information, legitimacy, and logistical reach, which in turn support sustained health and education outcomes.

\subsection{Dynamic persistence, cohort mechanisms, and heterogeneous long-run effects}
The long-run evolution of health and human-capital outcomes following a major reform is shaped by dynamic persistence and cohort-based accumulation. Early-life health improvements and better schooling conditions for one cohort can have spillovers on subsequent cohorts: more educated parents invest more in their children; higher incomes expand the fiscal base; stronger state capacity lowers the cost of maintaining and upgrading systems. Cohorts born just after the reform benefit from improved survival and health but may still face transitional constraints in schooling infrastructure; later cohorts experience both better health from birth and the full benefits of an expanded education system. These dynamics generate a self-reinforcing path in which initial reforms produce persistent and possibly widening differences. At the same time, different outcomes may respond differently to subsequent shocks. Life expectancy at birth is shaped not only by the quality of the health system but also by macroeconomic conditions, external shocks, and behavioral factors; economic crises, terms-of-trade collapses, or sharp fiscal contractions can attenuate or partially reverse some longevity gains. By contrast, average years of schooling, once literacy has been expanded, schooling norms have shifted, and institutions have been built, display stronger inertia and resilience. The conceptual framework thus suggests a pattern in which infant mortality responds quickly and persistently to improvements in health-system coverage and state capacity; life expectancy improves but remains exposed to subsequent macroeconomic shocks; and educational attainment exhibits a gradual but enduring divergence as healthier and better-prepared cohorts move through the schooling system.

\bibliographystyle{plainnat}
\bibliography{references}
\end{document}